\documentclass[a4paper, 12pt, oneside]{article}

\usepackage{amsmath, amsthm, amssymb, amsbsy}
\usepackage{authblk}
\usepackage{bm} 
\usepackage{braket}
\usepackage{caption}
\usepackage{cite}
\usepackage{enumitem} 
\usepackage{fancyhdr}
\usepackage[headheight=15pt]{geometry}
\usepackage{graphicx}
\usepackage[colorlinks=true, citecolor=blue, linkcolor=black]{hyperref}
\usepackage{mathptmx} 
\usepackage{tikz, tikz-cd} 
	\usetikzlibrary{decorations.pathmorphing}
\usepackage{verbatim}

\setitemize{itemsep=0.2\baselineskip,topsep=0pt, parsep=0pt,partopsep=0pt} 
\setenumerate{itemsep=0pt,topsep=0pt,parsep=0pt,partopsep=0pt} 

\newtheoremstyle{theorem}
			    {0.5\baselineskip} 
			    {0.5\baselineskip}	 
			    {\itshape} 
			    {0em} 
			    {\bfseries} 
			    {.}
			    {0.5em}
			    {}
\newtheoremstyle{definition}
			    {0.5\baselineskip} 
			    {0.5\baselineskip}	 
			    {} 
			    {0em} 
			    {\bfseries} 
			    {.}
			    {0.5em}
			    {}	
			      	    			    			    
\theoremstyle{theorem}
\newtheorem{theorem}{Theorem}

\theoremstyle{definition}
\newtheorem{definition}{Definition}

\newcommand{\CC}{\mathbb{C}}
\newcommand{\PP}{\mathbb{P}}
\newcommand{\RR}{\mathbb{R}}

\renewcommand{\aa}{\mathcal{A}}
\newcommand{\bb}{\mathcal{B}}
\newcommand{\bbr}{\mathcal{B}_{\mathbb{R}}}
\newcommand{\cc}{\mathcal{C}}
\newcommand{\cci}{\mathcal{C}^\infty}
\newcommand{\hh}{\mathcal{H}}
\newcommand{\kk}{\mathcal{K}}
\renewcommand{\ll}{\mathcal{L}}
\newcommand{\ph}{\mathbb{P}\mathcal{H}}
\newcommand{\pp}{\mathcal{P}}
\newcommand{\sh}{\mathbb{S}\mathcal{H}}
\renewcommand{\ss}{\mathcal{S}}
\newcommand{\uu}{\mathcal{U}}
\newcommand{\uur}{\mathcal{U}_{\mathbb{R}}}

\newcommand{\liet}{\mathfrak{t}}

\begin{document}
\pagestyle{fancy} 

\title{The Two-fold Role of Observables\\ in Classical and Quantum Kinematics}

\author[1,2]{Federico Zalamea%
\thanks{This work has received funding from the European Research Council under the European Community's Seventh Framework Programme (FP7/2007-2013 Grand Agreement n$^\circ$263523, ERC Project PhiloQuantumGravity).}}
\affil[1]{Laboratoire SPHERE - UMR 7219, Universit\'e Paris Diderot}
\affil[2]{Basic Research Community for Physics}

\date{}

\maketitle

\begin{abstract}

Observables have a dual nature in both classical and quantum kinematics: they are at the same time \emph{quantities}, allowing to separate states by means of their numerical values, and \emph{generators of transformations}, establishing relations between different states. In this work, we show how this two-fold role of observables constitutes a key feature in the conceptual analysis of classical and quantum kinematics, shedding a new light on the distinguishing feature of the quantum at the kinematical level. We first take a look at the algebraic description of both classical and quantum observables in terms of Jordan-Lie algebras and show how the two algebraic structures are the precise mathematical manifestation of the two-fold role of observables. Then, we turn to the geometric reformulation of quantum kinematics in terms of K\"ahler manifolds. A key achievement of this reformulation is to show that the two-fold role of observables is the constitutive ingredient defining what an observable is. Moreover, it points to the fact that, from the restricted point of view of the transformational role of observables, classical and quantum kinematics behave in exactly the same way. Finally, we present Landsman's general framework of Poisson spaces with transition probability, which highlights with unmatched clarity that the crucial difference between the two kinematics lies in the way the two roles of observables are related to each other. 

\end{abstract}

\newpage
\fancyhead[L]{F. Zalamea - \emph{The two-fold role of observables}}

\section{Introduction}
\label{intro}

In the contemporary usage, the `kinematical description' of a physical system has come to signify a characterization of all the \emph{states} accessible to the system and all the \emph{observables} which can be measured. These are the two fundamental notions of kinematics, and each is associated with different areas of mathematics: the set of all states is generally conceived as a \emph{space} and is hence described by means of \emph{geometric} structures; the set of all observables, on the other hand, is generally conceived as an \emph{algebra} and is accordingly described by means of algebraic structures. 

Of course, the notions of `state' and `observable' are closely related---much in the same way that, in mathematics, geometric and algebraic methods are. One first obvious such relation is the existence of a `numerical pairing' between states and observables. If we respectively denote by $\ss$ and $\aa$ the space of states and algebra of observables of a certain physical system, then the numerical pairing is a map:
\begin{eqnarray*}
\langle \cdot, \cdot \rangle: & \ss \times \aa &\longrightarrow \RR\\
& (\rho, F) &\longmapsto  \langle \rho, F \rangle.
\end{eqnarray*}
In the geometric formulation of classical kinematics, where the notion of state is primitive---it is the starting point from which the other notions are built---, this numerical pairing is seen as the \emph{definition} of an observable and is rather denoted by $F(\rho)$: observables are indeed defined as smooth real-valued functions over the space of states\cite{abraham1978, arnold1989}. On the contrary, in the algebraic formulation of quantum kinematics, the primitive notion is that of an observable and the numerical pairing is used instead to define states: the latter are considered to be linear (positive) functionals over the algebra of observables \cite{strocchi2008}. Accordingly, in the algebraic setting, the numerical pairing is denoted by $\rho(F)$. Formally, the transformation which allows to switch between these two points of view on the numerical pairing $\langle \rho, F\rangle$ (the geometric, where $\langle \rho, F\rangle=:F(\rho)$, and the algebraic, where $\langle \rho, F\rangle =:\rho(F)$) is called the \emph{Gelfand transform}\cite{gelfand1943, landsman1998}. 

As is well-known, one crucial difference between classical and quantum kinematics is the interpretation of the number $\langle \rho, F \rangle$. Whereas in the former the numerical pairing is interpreted as yielding the \emph{definite value} of the physical \emph{quantity} $F$ when the system is in the state $\rho$, in the latter the numerical pairing can only be interpreted \emph{statistically}---as the expectation value for the result of measuring the observable $F$ when the system is prepared in the state $\rho$. Because of this feature, the conception of observables as quantities having well-defined values at all times cannot be straightforwardly applied to the standard formulation of quantum kinematics. This difficulty has surely been one of the main sources of insatisfaction towards the quantum theory. In fact, one could argue that all hidden variable theories are (at least partially) motivated by the desire to reconcile quantum kinematics with such a conception. But, as the various results from von Neumann, Gleason, Bell, Kochen and Specker have shown, the clash between the standard quantum formalism and the interpretation of observables as quantities is irremediable\cite{neumann1955, gleason1957, bell1964, kochen1967}. 

Yet, although the numerical pairing and the associated conception of observables-as-quantities has dominated much of the attention, both classical and quantum observables play another important role in relation to states: they generate transformations on the space of states. The progressive disclosure of the intimate link between observables and transformations is, in my opinion, one of the most important conceptual insights that the 20th century brought to the foundations of kinematics.  A much celebrated result pointing in this direction is of course Noether's first theorem, which relates the existence of symmetries to the existence of conserved quantities\cite{kosmann2011}. For some particular observables, this relation is now included in the folklore of theoretical physics---for instance, by \emph{defining} linear momentum, angular momentum and the Hamiltonian as the generators of space translations, space rotations and time evolution respectively\cite{townsend2000}. This notwithstanding, the idea that a systematic relation between observables and transformations may constitute a key feature in the conceptual analysis of classical and quantum kinematics has remained somewhat dormant, despite some attempts to draw more attention to it\cite{alfsen2001, catren2008, guillemin2006, grgin1974, landsman1994}. 

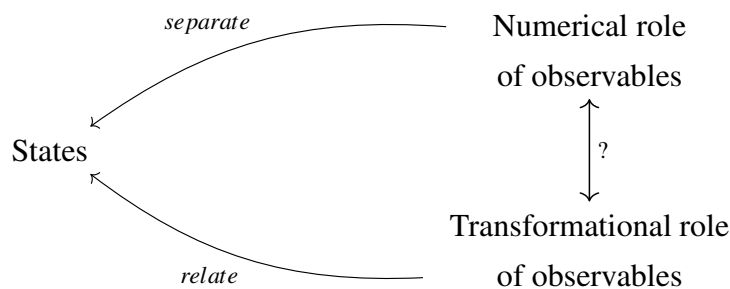
\begin{figure}[ht!]
	$$\begin{tikzcd}[column sep=large, row sep=small]
		&& \parbox{8em}{\centering \normalsize Numerical role\\ of observables}
			\ar[leftrightarrow, dd, "\Huge{?}", line width = 0.2mm] \\
		\parbox{8em}{\centering \normalsize States}
			\ar[leftarrow, rru, bend left=20, "separate", line width = 0.15mm]\\
		&& \parbox{9.45em}{\centering \normalsize Transformational role of observables}
			\ar[llu, bend left = 20, "relate", line width = 0.15mm]
	\end{tikzcd}$$
	\caption{The fundamental conceptual triad of kinematics.}
	\label{fig:triad}
\end{figure}

The goal of this paper is to insist on the usefulness of investigating the conceptual structure of both classical and quantum kinematics through the looking glass of the two-fold role of observables. Rather than considering ``states'' and ``observables'' as the two fundamental notions, we will henceforth distinguish observables-as-quantities and observables-as-transformations and consider what we call the ``fundamental conceptual triad of Kinematics" (\autoref{fig:triad}). Through their numerical role, observables allow to distinguish, to \emph{separate} different points of the space of states; on the other hand, when viewed as the generators of transformations on the space of states, they instead allow to \emph{relate} different states. Understanding precisely in which manner these two different roles are articulated to give a consistent account of the notion of ``observable'' will be the key question of our analysis. We will explain in detail how the two-fold role of observables is manifest in the mathematical structures used to describe the space of states and the algebra of observables of classical and quantum systems, and we will use this common feature to shed a new light on the fundamental traits distinguishing the Quantum from the Classical. As it will be shown, quantum kinematics can be characterized by a certain compatibility condition between the numerical and transformational roles of observables.

But before entering into the details, let us first briefly sketch the content of the paper. In \autoref{standard} we will review the standard formulations of classical and quantum kinematics, where the first is casted in the language of symplectic geometry and the second in the language of Hilbert spaces. In both cases, the algebra of observables has the structure of a Jordan-Lie algebra. This is a real algebra equipped with two structures, a commutative Jordan product and an anti-commutative Lie product, which respectively govern the numerical and transformational roles of observables. From this point of view, the only difference between the classical and quantum algebras of observables lies in the associativity or non-associativity of the Jordan product, but it is a priori unclear what this means.  

In \autoref{geometric}, we move on to discuss the geometric formulation of quantum kinematics, which stresses the role of the geometric structures inherent to any \emph{projective} Hilbert space. The Jordan-Lie structures of the algebra of observables are mirrored by two geometric structures on the quantum space of states: a symplectic and a Riemannian structure. A key achievement of this reformulation is to show that the two-fold role is the constitutive ingredient defining what an observable is. Moreover, it points to the fact that, from the restricted point of view of the transformational role of observables, classical and quantum kinematics behave in exactly the same way. 

However, a satisfactory comparison of the classical and quantum Jordan structures remains somewhat elusive at this stage, mainly because the language of K\"ahler manifolds fails to provide a unifying language for describing both kinematics. Thus, in \autoref{landsman} we finally turn to Landsman's proof that any state space can be described as a Poisson space with transition probability. As we will show, this framework highlights with unmatched clarity that the crucial difference between classical and quantum kinematics lies in the way the two roles of observables are related to each other.


\section{Standard formulation of kinematics}
\label{standard}

\subsection{Standard classical kinematics}
\label{standardC}

\begin{flushright}
\begin{minipage}{0.74\textwidth}
{\small \emph{Symplectic geometry has become the framework \emph{per se} of mechanics, up to the point one may claim today that these two theories are the same. Symplectic geometry is not the language of mechanics, it is its essence and matter.}\cite{iglesias2014}}
\begin{flushright}
{\small P. Iglesias-Zemmour}
\end{flushright}
\end{minipage}
\end{flushright}

Classical Hamiltonian mechanics is casted in the language of symplectic geometry\cite{souriau1970, chernoff1974, abraham1978, arnold1989, puta1993, marsden1999}. In this formulation, the starting point is the classical space of states of the system $\ss^C$, which is identified with a finite-dimensional symplectic manifold. 

\begin{definition}
\label{def:symp}
A finite-dimensional \emph{symplectic manifold} is a differentiable manifold $S$ equipped with one additional structure: a two-form $\omega \in \Omega^2(S)$, called the \emph{symplectic form}, which is closed and non-degenerate. This means:
\begin{enumerate}[label=\roman*), leftmargin=1cm]
\item  $\omega$ is an anti-symmetric section of $T^*S \otimes T^*S$,
\item $d\omega = 0$,
\item $\omega$, seen as a map from $TS$ to $T^*S$, is an isomorphism.\index{symplectic! manifold|textbf}\index{symplectic! 2-form|textbf}
\end{enumerate}
\end{definition}

The group $T^C$ of classical \emph{global state transformations} is the group of symplectomorphisms $\text{Aut}(S) = Symp (S)$\footnote{%
Sometimes, these transformations are also called \emph{canonical transformations}.
}. 
It is the subgroup of diffeomorphisms $\phi : S \longrightarrow S$ leaving invariant the symplectic 2-form: $\phi^*\omega=\omega$, where $\phi^*\omega$ is the pull-back of the symplectic form\footnote{%
A diffeomorphism $\phi: S \longrightarrow S$ induces a map $\Phi: \cci(S, \RR) \longrightarrow \cci(S, \RR)$ defined by: 
$$\forall f \in \cci(S, \RR), (\Phi f)(p) = f(\phi(p)).$$ 
This in turn allows one to define the \emph{push-forward} $\phi_*$ of vector fields and the \emph{pull-back} $\phi^*$ of $n$-forms by: 
\begin{align*}
\forall v \in \Gamma(TS),& \:\: (\phi_*v)[f]:= v[\Phi f],\\
\forall \alpha \in \Omega^n(S),& \:\: (\phi^*\alpha)(v_1, \ldots, v_n):= \alpha(\phi_*v_1, \ldots,\phi_*v_n ).
\end{align*}
\vspace*{-1.5em}
}.

The Lie algebra $\liet^C$ of classical \emph{infinitesimal state transformations} is the Lie algebra associated to the group of global transformations. It is the Lie algebra $\Gamma(TS)_\omega$ of vector fields leaving invariant the symplectic 2-form: $\Gamma(TS)_\omega = \{ v \in \Gamma(TS) \:|\: \ll_v \omega = 0\}$ where $\ll$ denotes the Lie derivative\footnote{%
For a given two-form $\alpha \in \Omega^2(S)$, the Lie derivative with respect to the vector field $v \in \Gamma(TS)$ is given by the so-called ``Cartan's magic formula'': $\ll_v \alpha = (\iota_v d + d\iota_v)\alpha$, where $\iota_v \alpha := \alpha(v, \cdot) \in \Omega^1(S)$.
}.

Finally, classical observables are defined as smooth real-valued functions over the space of states. The algebra of observables $\cci(S, \RR)$ has the structure of a Poisson algebra:

\begin{definition}
\label{def:Poiss}
A \emph{Poisson algebra} is a real (usually infinite-dimensional) vector space $\aa^C$ equipped with two additional structures: a Jordan product $\bullet$ and a Lie product $\star$ such that:
\begin{enumerate}[label=\roman*), leftmargin=1cm]
\item $\bullet$ is a bilinear symmetric product, \label{Psym}
\item $\star$ is a bilinear anti-symmetric product,\label{Pasym}
\item $\star$ satisfies the Jacobi identity: $f\star (g\star h) + g\star (h\star f) + h\star (f\star g) = 0$,\label{PJac}
\item $\star$  satisfies the Leibniz rule with respect to $\bullet$: $f\star (g \bullet h) = (f \star g)\bullet h + g \bullet (f \star h)$,\label{PLeib}
\item $\bullet$ is associative. \label{Passoc}
\end{enumerate}
The Lie product of a Poisson algebra is very often called the \emph{Poisson bracket} and denoted by $\{\cdot, \cdot\}$.
\end{definition}

In this case, the commutative and associative Jordan product $\bullet$ is simply the point-wise multiplication of functions. The Lie product, on the other hand, is defined in terms of the symplectic structure by:
\begin{equation}
\label{def:pb}
\forall f,g \in \cci(S, \RR), \: \{f,g\}=f \star g:=\omega(df^\sharp, dg^\sharp),
\end{equation}
where $df^\sharp:=\flat^{-1}(df)$ and $\flat$ denotes the so-called musical vector bundle isomorphism defined by 
\begin{align*}
\flat : T_pS \xrightarrow{\:\: \sim \:\:}& \:T_p^*S\\
v \longmapsto &\: \omega_p(v, \cdot).
\end{align*}
The fact that $\omega$ is a 2-form implies the anti-commutativity of the product thus defined, whereas the Jacobi identity follows from the closedness of the symplectic form. 

Let us make a series of comments on the definition of the Poisson algebra of classical observables in order to motivate the terminology and explain  the relation between the algebraic structures and the two-fold role of observables in classical kinematics. 

First, axioms \ref{Psym} and \ref{Passoc} turn $(\aa^C, \bullet)$ into a Jordan algebra\footnote{%
A \emph{real Jordan algebra} $(\aa, \bullet)$ is a commutative algebra such that, moreover, $F \bullet (G \bullet F^2) =( F \bullet G) \bullet F^2$ for all $F, G \in \aa$. This concept was introduced by the German theoretical physicist Pascual Jordan in 1933\cite{jordan1933}.  
}. 
Moreover, notice how the very definition of the Jordan product of two classical observables involves solely their \emph{numerical} role: $f\bullet g$ is defined as the observable whose \emph{value} at each state is the product of the \emph{values} of the observables $f$ and $g$ at the the same state. Conversely, the set $spec(f) \subset \RR$ of values of the observable $f$ is in fact completely determined by its position within the Jordan algebra $(\aa^C, \bullet)$ \cite{landsman1994}. Indeed, it may be defined as 
\begin{equation*}
spec(f):=\big\{\alpha \in \RR \:\big|\: \nexists g \in (\aa^C, \bullet) \text{ such that } (f-\alpha1) \bullet g = 1\big\},
\end{equation*}
in exact analogy with the definition of the spectrum of a linear operator\footnote{%
Recall that, given a vector space $V$ and a linear operator $A$ acting on $V$, the spectrum of $A$ is defined as 
\begin{equation*}
spec(A):=\big\{\alpha \in \RR \:\big|\: (A-\alpha\, \text{Id}_V) \text{ is not invertible}\big\}.
\end{equation*}
\vspace*{-0.5em}
}. 
In this sense, \emph{the Jordan structure of the algebra of classical observables completely encodes their numerical role}. 

Similarly, axioms \ref{Pasym} and \ref{PJac} turn $(\aa^C, \star)$ into a Lie algebra. Only axiom \ref{PLeib} establishes a relation between the otherwise unrelated Jordan and Lie structures of the algebra of classical observables. Given an observable $f \in \aa^C$, consider the linear operator $v_f$ whose action on any element $g\in \aa^C$ is defined by $v_f(g):=f \star g$. The Leibniz rule states that the linear operator $v_f$ is in fact a derivation on the Jordan algebra $(\aa^C, \bullet)$. Now, derivation on an algebra of smooth functions over a manifold are nothing but vector fields:
$$Der(\cci(S, \RR), \bullet) = \Gamma(TS),$$
and it is easy to show that the derivative operator $v_f$ leaves the symplectic form invariant. Hence, the Leibniz rule guarantees the existence of a map
\begin{equation}
\label{map:cot}
v_{-}: \aa^C \longrightarrow \liet^C
\end{equation}
that, to any classical observable $f$ associates an infinitesimal state transformation $v_f$. The vector field $v_f$ is more commonly called the \emph{Hamiltonian vector field} associated to $f$, and $v_{-}$ the \emph{Hamiltonian map}. It is the technical tool that captures the transformational role of classical observables. In particular, the susbset $\liet_\aa^C$ of Hamiltonian vector fields represents the set of infinitesimal transformations arising from classical observables. 

From this point of view, the Jacobi identity is the requirement that this map be a morphism of Lie algebras: 
$\begin{tikzcd}
(\aa^C, \star) \ar[r, "v_{-}"] & \liet^C.
\end{tikzcd}$
Indeed, axiom \ref{PJac} may be rewritten as:
\begin{eqnarray*}
v_{f\star g}(h) = v_f \circ v_g(h) - v_g \circ v_f(h) =:[v_f, v_g](h).\footnotemark
\end{eqnarray*}
\footnotetext{%
$v_f$ and $v_g$ being linear operators on the real vector space $\aa^C$, one can consider their composition $v_f \circ v_g$. This fails to be a derivative operator (and hence a vector field), but the commutator $v_f \circ v_g-v_g \circ v_f$ is again a derivative operator.
}%
Moreover, since the kernel of the map $v_{-}$ is the set of constant functions\footnote{%
Here, we suppose that the space of state is a simply connected manifold. In the general case, the kernel of $v_{-}$ is the center of $(\aa^C, \star)$, that is, the set of \emph{locally} constant functions.
}, 
we then have the isomorphism of Lie algebras
\begin{equation}
\label{iso:cot}
(\aa^C/ \RR, \star) \simeq \liet^C_\aa.
\end{equation}
In other words, ``up to a constant'', \emph{the transformational role of classical observables is found by simply forgetting the Jordan product and focusing on the Lie structure}.

\begin{figure}[ht!]
$${\small\begin{tikzcd}[row sep=small]
	 &&&& \parbox{7em}{\centering \normalsize \textbf{numerical role}\\ \textbf{of observables}} \ar[dd, "v_{-}", Rightarrow]
	 	& \parbox{7em}{\centering \emph{Jordan structure}\\ (commutative, associative)}\\
		&\boxed{\normalsize\textsc{states}} \ar[rrru, bend left=20, Rightarrow, line width = 0.23mm]\\
	\parbox{4em}{\centering \emph{symplectic structure}}
	&&&& \parbox{9em}{\centering \normalsize transformational role of observables} \ar[lllu, bend left = 20, dashed]
	 	& \parbox{8em}{\centering \emph{Lie structure}\\ (non-commutative, non-associative)}
\end{tikzcd}}$$
\caption{Relation between the conceptual and mathematical structures in the standard geometric formulation of classical kinematics. Double arrows indicate conceptual priority. Thus, in this formulation, the notion of state (boxed) is the most primitive one and the numerical role (in bold typeface) appears as the defining feature of classical observables. In a third stage, the geometric symplectic structure is used to define their transformational role and the associated algebraic Lie structure. The dashed arrow simply indicates that, through this construction, observables act on the space of states.
}
\label{fig:SC}
\end{figure}
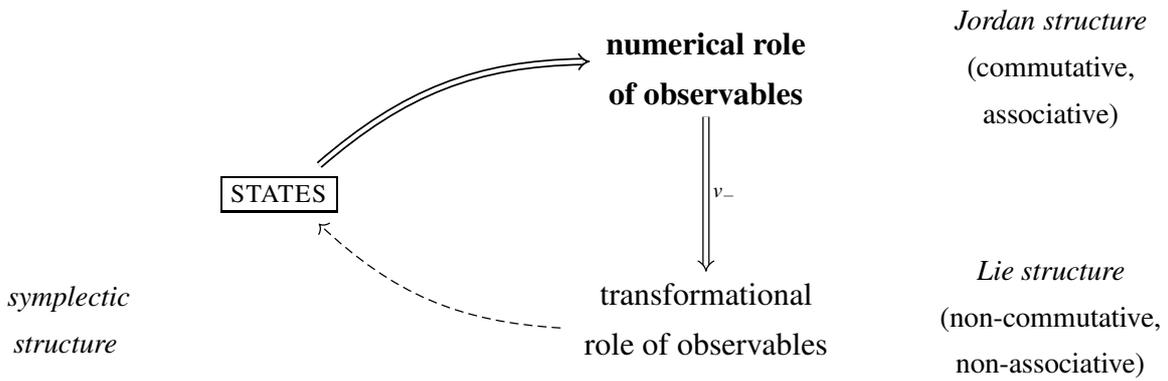

To sum up, the following picture emerges (see \autoref{fig:SC}): In the standard geometric formulation of classical kinematics, the primitive notion from which one constructs all the others is the notion of `state'. Classical observables are defined by their numerical role, which yields the \emph{commutative} Jordan algebra $(\cci(S, \RR), \bullet)$. Thus, classical observables are primarily seen as \emph{quantities}. Their transformational role, on the other hand, appears only as a secondary feature, defined in a subsequent stage. It is defined through the addition of a \emph{non-commutative} algebraic structure, the Lie product or Poisson bracket, induced by the geometric symplectic structure present on the classical space of states.
\vspace*{-1.5em}

\subsection{Standard quantum kinematics}
\label{standardQ}

In the standard formulation of quantum kinematics, the starting point is an abstract Hilbert space $\hh$, usually infinite-dimensional. In order to facilitate the comparison with the classical case, we will again identify the mathematical structures used to describe the four fundamental structures: the space of quantum states $\ss^Q$, the group of quantum global state transformations $T^Q$, the Lie algebra of quantum infinitesimal state transformations $\liet^Q$ and the algebra of quantum observables $\aa^Q$. 

\begin{definition}
\label{def:Hilb}
A complex \emph{Hilbert space} is a complex vector space $\hh$ equipped with one additional structure: a Hermitian, positive-definite map $\langle \cdot, \cdot \rangle: \hh \times \hh \longrightarrow \CC$ such that the associated metric $d: \hh \times \hh \longrightarrow \RR^+$ defined as $d(\psi, \varphi) := \langle \varphi - \psi, \varphi - \psi\rangle$ turns  $(\hh, d)$ into a complete metric space.
\end{definition}

For a quantum system described by $\hh$, states are given by rays of the Hilbert space---that is, by one-dimensional subspaces of $\hh$. 

The group $T^Q$ of \emph{global state transformations} is the group $Aut(\hh)= U(\hh)$ of unitary operators. It is the subgroup of linear operators $U: \hh \longrightarrow \hh$ such that $U^*= U^{-1}$. The Lie algebra $\liet^Q$ of \emph{infinitesimal state transformations} is the Lie algebra $(\bb_{i\RR}, [\cdot, \cdot])$ of bounded anti-self-adjoint operators\footnote{%
For finite-dimensional Hilbert spaces, this is clear: any operator $A \in \bb_{i\RR}$ defines a one-parameter group of unitary operators through exponentiation: $e^{tA} \in U(\hh),\: t \in \RR$. The situation is more delicate in the infinite-dimensional case for two reasons. First, $U(\hh)$ is not a Lie group (it is infinite-dimensional) and thus the notion of an associated Lie algebra is problematic. However, by Stone's theorem we know there is a one-to-one correspondence between anti-self-adjoint operators and continuous one-parameter unitary groups. In this sense, one is still allowed to claim that anti-self-adjoint operators are the generators of unitary transformations.  The second problem is that, without further restrictions, anti-self-adjoint operators do not form a Lie algebra (in fact, they do not even form a vector space). This is the reason why we restrict attention here to \emph{bounded} anti-self-adjoint operators.   For a precise mathematical treatment of these issues, see \cite{abraham1988}.
}. 

Finally, a quantum observable is described by a bounded self-adjoint operator. The algebra of observables $\bbr(\hh)$ has the structure of a non-associative Jordan-Lie algebra:
\begin{definition}
\label{def:naJL}
A \emph{non-associative Jordan-Lie algebra} is a real vector space $\aa^Q$ (usually infinite-dimensional) equipped with two additional structures: a Jordan product $\bullet$ and a Lie product $\star$ such that
\begin{enumerate}[label=\roman*), leftmargin=1cm]
\item $\bullet$ is a bilinear symmetric product, \label{naJLsym}
\item $\star$ is a bilinear anti-symmetric product,\label{naJLasym}
\item $\star$ satisfies the Jacobi identity: $F\star (G\star H) + G\star (H\star F) + H\star (F\star G) = 0$,\label{naJLJac}
\item \label{naJLLeib}$\star$  satisfies the Leibniz rule with respect to $\bullet$: $$F\star (G \bullet H) = (F \star G)\bullet H + G \bullet (F \star H),$$
\item \label{naJLassoc} $\bullet$ and $\star$ satisfy the associator rule: $(F \bullet G) \bullet H - F \bullet (G\bullet H) = (F \star H) \star G.$ 
\end{enumerate}
\end{definition}

In this case, both the Jordan and Lie products are related to the composition of operators $\circ$, by means of the anti-commutator and ($i$-times) the commutator respectively:
\begin{eqnarray}
F\star G :=& \frac{i}{2}[F, G]_{~} = \frac{i}{2}(F \circ G-G \circ F)\:\:
\label{def:qjp}\\
F \bullet G :=& \frac{1}{2}[F, G]_+= \frac{1}{2}(F\circ G + G\circ F)\footnotemark.
\label{def:qlp}
\end{eqnarray}
\footnotetext{
It is important to stress that these are the two natural structures present on the set of \emph{self-adjoint} operators. For example, the composition of operators is not a well-defined operation on this set (the composition of self-adjoint operators is not self-adjoint). One should also notice that the Lie product on bounded self-adjoint operators is \emph{not} the commutator: the multiplication by the complex number $i$ in the definition is a necessary one. This is because the commutator of two self-adjoint operators yields an \emph{anti}-self-adjoint operator. On the other hand, the two factors $\frac{1}{2}$ are only a convenient normalization in order to obtain the associator rule as written in axiom \ref{naJLassoc} but other choices are possible. For instance, another normalization is $F\star G := \frac{i}{\hbar}[F, G]$, which forces $\kappa = \frac{\hbar^2}{4}$ (cf. \autoref{def:JL}), but allows one to write the canonical commutation relations between position and momentum operators as $P \star X = 1$\index{canonical! commutation relations}.
}%

As it was the case in classical kinematics, also in quantum kinematics can the two natural algebraic structures present on the set of quantum observables be seen as the manifestation of the two-fold role of observables. This time, however, the transformational role of quantum observables is much easier to perceive. Indeed, the quantum analogue of the classical map \eqref{map:cot} is here simply defined as 
\begin{align}
V_{-}: \aa^Q &\longrightarrow \liet^Q \label{map:qot}\\
F &\longmapsto iF\nonumber.
\end{align}
In other words, given a quantum observable $F$, the associated generator of state transformations is just the anti-self-adjoint operator obtained through multiplication by $i$. The map $V_{-}$ is obviously an isomorphism of Lie algebras:
\begin{equation}
(\aa^Q, \star) \simeq \liet^Q,
\label{iso:qot}
\end{equation}
which should be compared with its classical analogue \eqref{iso:cot}. Again, this means that \emph{considering quantum observables solely in their transformational role---that is, ignoring their numerical role---corresponds exactly to focusing only on the Lie structure and forgetting the second algebraic structure} (here, the Jordan product). 

Mathematically, the above statement is certainly trivial. But this triviality points to the fact that, whereas in classical kinematics there was an emphasis on the numerical role of observables, quantum kinematics, at least in the standard Hilbert space formulation, presents the reverse situation: \emph{quantum observables are defined by their role as generators of state transformations} and the reading of their numerical role is more involved. From this perspective, it is not surprising that specific quantum observables are sometimes explicitly defined through their transformational role\cite{townsend2000}. 

On the other hand, the set of possible values of a quantum observable $F$ is completely determined by the Jordan structure. A simple way of seeing this is to recall that the spectrum of $F$ coincides with the `Gelfand spectrum' of the $C^*$-algebra $(C^*(F), \circ)$ generated by $F$\footnote{%
Given a unital commutative $C^*$-algebra $\uu$, its \emph{Gelfand spectrum} $spec_G(\uu)$ is the set of all positive linear functionals $\rho: \uu \longrightarrow \CC$ such that $\rho(\mathbb{I})=1$\cite{landsman1998}. The fact that $spec_G(C^*(F))$ is isomorphic to the spectrum of $F$ (in the usual sense) justifies the use of the word ``spectrum" in Gelfand's theory\cite{cartier2008}.  
}.
But when $F$ is self-adjoint, this is a \emph{commutative} subalgebra of $\bb(\hh)$\footnote{%
Indeed, $C^*(F)$ consists of all polynomials in $F$ and $F^*$. Then, $C^*(F)$ will be non-commutative if and only if $[F, F^*]\neq 0$.
}, 
in which case the composition $\circ$ and the anti-commutator $\bullet$ on $C^*(F)$ are the same operation. Therefore, as it was the case for the Classical, the Jordan structure encodes all the information of the numerical role of quantum observables.

At this point, it is worth drawing the analogue of \autoref{fig:SC} for the standard formulation of quantum kinematics. Now, the primitive mathematical structure from which all others are constructed is that of a Hilbert space $\hh$. But $\hh$ does not correspond neither to the space of states (given by $\ph$) nor to the algebra of observables (given by $\bbr(\hh)$). Thus, one can no longer say which, among ``states" and ``observables", is the primitive notion of the formulation. The two roles of observables, on the other hand, are not on the same footing: observables are here defined by their transformational role (as operators on states) and their numerical role only comes in a second step (as expectation-values of operators). The conceptual priority between the two roles of observables is therefore reversed with respect to the classical case discussed in the previous section. The conceptual diagram is summarised in \autoref{fig:SQ} below.

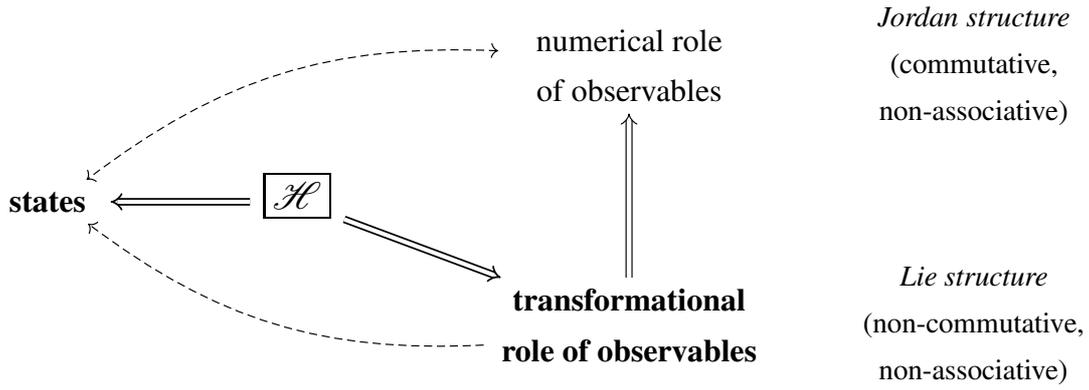
\begin{figure}[ht!]
{\small$$\begin{tikzcd}[row sep=small]
	 &&&& \parbox{8em}{\centering \normalsize numerical role\\ of observables} \ar[dd, Leftarrow] \ar[lllld, bend right = 20, dashed, leftrightarrow]
	 	& \parbox{7em}{\centering \emph{Jordan structure}\\ (commutative, non-associative)}\\
		\parbox{3em}{\normalsize \textbf{states}} & & 
		{\large\boxed{\hh}}  \ar[ll, Rightarrow, line width = 0.23mm]
		\ar[rrd, Rightarrow, line width = 0.23mm]
		\\
	&&&& \parbox{9em}{\centering \normalsize \textbf{transformational}\\\textbf{role of observables}} \ar[llllu, bend left = 20, dashed]
	 	& \parbox{8em}{\centering \emph{Lie structure}\\ (non-commutative, non-associative)}
\end{tikzcd}$$}
\vspace*{-1em}
\caption{Relation between the conceptual and mathematical structures in the standard  formulation of quantum kinematics in terms of Hilbert spaces. Again, double arrows indicate conceptual priority. Here, the primitive mathematical structure is that of a Hilbert space $\hh$ (boxed), which allows to define both what states and observables are. The latter are defined by their transformational role, the numerical role being in this formulation the secondary one. Here again, the two-fold role of observables is captured by the presence of the two algebraic structures, but their geometric interpretation is missing.
}
\label{fig:SQ}
\end{figure}

\subsection{Classical/Quantum: from associative to non-associative Jordan-Lie algebras}
\label{assoc/nassoc}

The striking similarity just brought to light between the classical and quantum algebras of observables motivates the definition of a general (not necessarily non-associative) Jordan-Lie algebra, which encapsulates both the classical and the quantum cases\cite{landsman1998}:
\begin{definition}
\label{def:JL}
A \emph{general Jordan-Lie algebra} is a real vector space $\aa$ equipped with two additional structures: a Jordan product $\bullet$ and a Lie product $\star$ such that
\begin{enumerate}[label=\roman*), leftmargin=1cm]
	\item $\bullet$ is a bilinear symmetric product, \label{JLsym}
	\item $\star$ is a bilinear anti-symmetric product,\label{JLasym}
	\item $\star$ satisfies the Jacobi identity: $F\star (G\star H) + G\star (H\star F) + H\star (F\star G) = 0$,\label{JLJac}
	\item \label{JLLeib}$\star$ satisfies the Leibniz rule with respect to $\bullet$: $$F\star (G \bullet H) = (F \star G)\bullet H + G \bullet (F \star H),$$
	\item \label{JLassoc} $\bullet$ and $\star$ satisfy the associator rule: $$\exists \kappa \in \RR, (F \bullet G) \bullet H - F \bullet (G\bullet H) = \kappa^2(F \star H) \star G.$$
\end{enumerate}
\end{definition} 

Only the last axiom differentiates the classical and quantum algebras of observables. When $\kappa=0$, the Jordan product is associative and one gets the definition of a Poisson algebra describing classical observables. When $\kappa=1$, one gets the previous definition for the algebra of quantum observables with a non-associative Jordan product. In fact, whenever $\kappa \neq 0$, one may always rescale the Lie product so as to yield $\kappa = 1$. Therefore, the world of Jordan-Lie algebras\index{Jordan-Lie algebras} is sharply divided into the sole cases of $\kappa = 0$ (corresponding to classical mechanics) and $\kappa = 1$  (corresponding to quantum mechanics). In this precise sense, one can say that \emph{the transition from classical observables to quantum observables is the transition from associative Jordan-Lie algebras to non-associative Jordan-Lie algebras}.

Characterizing the algebraic difference between Classical and Quantum in terms of the associativity or non-associativity of the Jordan product may come as a surprise to the reader more familiar with the widespread conception ``classical/quantum = commutative/non-\-com\-mutative''. However, it is by no means clear how to render precise this heuristic equation. As it is unfortunately often the case, the equation can be taken to mean that in Classical Kinematics one always has $fg-gf=0$ whereas in Quantum Kinematics one has in general $[F,G]\neq 0$. But this point of view adopts a wrong analogy between the two kinematics. Indeed, instead of comparing either the full algebras of observables (with both the Jordan and Lie structures), the two commutative algebras of observables-as-quantities (with only the Jordan structure) or else the two non-commutative algebras of observables-as-transformations (with only the Lie structure), this point of view on the Classical/Quantum transition compares point-wise multiplications of functions with the commutator of operators---that is, the Jordan structure of classical observables with the Lie structure of quantum observables...\footnote{%
This confusion was there since the very beginning of Quantum Mechanics. For example, in their second paper of 1926, Born, Heisenberg and Jordan write: 
\begin{quote}
We introduce the following basic quantum-mechanical relation: $\bm{pq} - \bm{qp} = \frac{h}{2\pi i} \bm{1}$. [...] One can see from [this equation] that in the limit $h=0$ the new theory would converge to classical theory, as is physically required. \cite[327]{born1967a}
\end{quote}
It is clear that they were comparing the commutator in quantum mechanics with point-wise multiplication in Classical Mechanics (despite the fact that, by the time of the second quoted paper, Dirac had already shown in \cite{dirac1925} that the quantum commutator should be compared to the classical Poisson bracket). 
}

Another widespread, and mathematically more sophisticated, point of view from which one could try to give meaning to the commutativity/non-commutativity idea is that of $C^*$-algebras\footnote{
A $C^*$-algebra $(\uu, \circ, *, \|\cdot\|)$ is a complex associative algebra $(\uu, \circ)$ equipped with an involution $^*$ and a norm $\| \cdot \|$ such that: \emph{i)} $(\uu, \| \cdot \|)$ is a complex Banach space, \emph{ii)} $\forall A, B \in \uu, \| A \circ B \| \leq \| A \|\| B \|$, and \emph{iii)} $\forall A \in \uu, \| A^* \circ A \| = \| A \|^2$.
}%
: classical observables would be described by commutative $C^*$-algebras and quantum observables by non-commutative ones. However, although partially correct, this viewpoint  fails to capture the whole situation. It is indeed true that the full algebra of quantum observables can equivalently be described as real non-associative Jordan-Lie-Banach algebra or as a complex non-commutative $C^*$-algebra. But, on the other hand, it is simply false that the full \emph{Poisson} algebra of classical observables can be described as a commutative $C^*$-algebra. In fact, describing the set of classical observables as a commutative $C^*$-algebra is equivalent to completely ignoring the classical Poisson structure!\footnote{
To be more precise: given any $C^*$-algebra $(\uu, \circ)$, its real part $\uur:=\big\{A \in \uu \big| A=A^*\big\}$ equipped with the operations $A\bullet B= \frac{1}{2}(A\circ B+ B\circ A)$ and $A\star B= \frac{i}{2}(A\circ B- B\circ A)$ is a Jordan-Lie-Banach algebra. Conversely, given any real JLB-algebra $(\uur, \bullet, \star)$, its complexification $(\uur)_\CC$ can be turned into a $C^*$-algebra by defining the operation $A \circ B := A \bullet B -i A \star B$. In this sense, $C^*$-algebras are equivalent to JLB-algebras. Moreover, a $C^*$-algebra $\uu$ is commutative if and only if the associated JLB-algebra $\uur$ is associative. However, not all Jordan-Lie algebras can be equipped with a norm so that they become JLB-algebras, and, in fact, no non-trivial Poisson algebra can be normed in such a way that the bracket is defined in the norm-completion of the algebra. Therefore, non-trivial Poisson algebras do not fall under the theory of $C^*$-algebras. For details, see \cite[Chapter I.1]{landsman1998}.
}

Thus, when it comes to characterizing the algebraic difference between both kinematics, it seems better to abandon the commutativity/non-commutativity picture and adopt the associativity/non-associativity opposition furnished by the language of Jordan-Lie algebras. The main question becomes then to understand the meaning of the associator rule, crucial in distinguishing both kinematics. When $\kappa\neq 0$, as in the quantum case, a new relation between the Jordan and Lie structures of the algebra of observables is introduced. Therefore, one would expect that the precise way in which the two roles of observables are intertwined differs in both theories. In this regard, it is interesting to note, as it has been done in \cite{catren2014a}, that two of the most notable features of the Quantum---namely, the eigenstate-eigenvalue link and the existence of a condition for observables to be compatible---may be reformulated as conditions relating the numerical and transformational roles:
\begin{enumerate}[leftmargin=1cm, label=\arabic*.]
\item \emph{Eigenstate-eigenvalue link}. In the standard formulation of quantum kinematics, an observable $F$ has a definite value only when the state of the system is described by an eigenstate of the operator $F$: $F|\rho\rangle= \tilde F(\rho) |\rho \rangle$. If we denote by $[\rho]$ the ray in $\hh$ describing the state of the system, this condition may be reformulated as:  \emph{the observable $F$ has a definite \textbf{value} on the state $[\rho]$ if and only if the state is left invariant by the \textbf{transformations} associated to the observable: $e^{tV_F}[\rho]= [\rho]$}.

\item \emph{Compatibility of observables}. Two observables $F$ and $G$ are said to be compatible if, for any $\epsilon \in \RR^+$, it is possible to prepare the system in a state $\rho$ such that $\Delta_\rho(F) + \Delta_\rho(G) < \epsilon$, where $\Delta_\rho(F)$ denotes the uncertainty of $F$\footnote{%
Because of Heisenberg's famous uncertainty relations, the definition of compatible observables is perhaps more often stated in terms of the product $\Delta_\rho(F) \Delta_\rho(G)$ rather than the sum. However, as Strocchi points out in \cite{strocchi2008}, this is wrong since for any two \emph{bounded} operators, one has $\inf\limits_\rho(\Delta_\rho(F) \Delta_\rho(G)) =0$. 
}. 
Thus, the compatibility of two observables is a notion that applies to their numerical role. However, as it is well-known, two observables are compatible if and only if their Lie product vanishes: $F \star G = \frac{i}{2}[F,G]=0$. This may be reformulated as: \emph{the \textbf{values} of two quantum observables $F$ and $G$ are compatible if and only if $F$ is invariant under the \textbf{transformations} generated by $G$ (or vice versa)}. 
\end{enumerate}

The possibility of these reformulations hints to the idea that indeed the quantum is characterized by a particular interplay between the two-fold role of observables. But to confirm this, it is necessary to have a deeper understanding of the Jordan and Lie structures governing the algebra of observables.

\vspace*{-0.5em}
\section{The geometric formulation of quantum kinematics}
\label{geometric}

The comparison of the standard formulations of both kinematics brings out a striking structural similarity between the algebras of classical and quantum observables. They are both equipped with two products---one commutative and one anti-commutative---whose existence may be seen as a manifestation of the fundamental two-fold role of the observables of a physical system. On the other hand, the classical and quantum descriptions of the space of states seem at first sight not to have any points in common. One could then be inclined to think that, although the non-associativity of the Jordan product has been spotted as the main algebraic difference between classical and quantum observables, the really crucial departure of the Quantum with respect to the Classical lies in the nature of the space of states. For in the dominant conception of quantum mechanics, the linearity of the space of states is concomitant of the \emph{superposition principle}\index{superposition principle}, which in turn is often regarded as one---or perhaps \emph{the}---fundamental feature of the theory, as Dirac asserts:
\vspace*{-0.5em}
\begin{quote}
{\small For this purpose [of building up quantum mechanics] a new set of accurate laws of nature is required. One of the most fundamental and the most drastic of these is the \emph{Principle of Superposition}.\cite{dirac1958}}
\end{quote}
                                                                                                                                                                                                                                                                                                                                                                                                                                                                                                                                                                                                                                                                                                                                                                                                                                                                                                                                                                                                                              
From this perspective, the apparently radical difference between the geometric space of classical states and the linear space of quantum states may be perceived as the natural---and almost necessary---manifestation of this ``drastic'' new feature of the Quantum. But in claiming so, one forgets a central point, which indicates this whole idea cannot be the end of the story: the ``true'' quantum space, in which points do represent states, is the projective Hilbert space $\PP\hh$, a genuine \emph{non-linear} manifold.

The principle of superposition\index{superposition principle} has certainly been a powerful idea, with a strong influence on the heuristics of the Quantum, and its link with the linearity of Hilbert spaces has been in my opinion one of the main reasons for the widespread use of the standard formalism. However, in the attempt to compare classical and quantum kinematics, due care should be taken to express both kinematics in as similar terms as possible. It becomes therefore natural to attempt a reformulation of the quantum situation in a language resembling the classical one---that is, to forget Hilbert spaces and to develop the quantum theory directly in terms of the intrinsic geometry of $\PP\hh$. 

The task of this reformulation is sometimes referred to as the ``\emph{geometric or delinearization program}''\index{delinearization program}. Its explicit goal is to reestablish the fruitful link, witnessed in the classical case, between the geometry of the space of states and the algebraic structures of observables. Some of the most important references are \cite{kibble1979, brody2001, ashtekar1997, schilling1996, cirelli1999, cirelli2003}\footnote{%
It is important to clearly distinguish the program of a geometric reformulation of quantum mechanics from the program of `geometric quantization' which we will not discuss here and is completely unrelated. The first aims at a reformulation of quantum mechanics which avoids Hilbert spaces. The second is geared towards an explicit construction of the quantum description of a system for which the classical description is given. But the resulting quantum description is still based on Hilbert spaces. What is `geometric' about geometric quantization is the means by which the Hilbert space is constructed: roughly, one starts with the symplectic manifold describing the classical system, considers a complex line bundle over it and defines the Hilbert space as a particular class of sections of this bundle. The program of geometric quantization was started by Jean-Marie Souriau and Bertram Kostant \cite{souriau1970, kostant1970}. A standard reference is \cite{woodhouse1991}. 
}. 
  
The central result upon which the whole geometric program is based is the fact the projective Hilbert space $\ph$ is a K\"ahler manifold.

\begin{definition}
A \emph{K\"ahler manifold} is a real manifold $M$ (possibly infinite-dimensional,
in which case $M$ should be taken to be a Banach manifold) 
equipped with two additional structures: a symplectic form $\omega$ and an integrable almost complex structure $J$ which is compatible with $\omega$. This means:
\begin{enumerate}[label=\roman*), leftmargin=1cm]
\item $J$ is a vector bundle isomorphism $J : TM \longrightarrow TM$ such that $J^2 =-1$,
\item For any point $p$ of $M$ and any two vectors $v,w$ of $T_pM$,
\begin{equation}
\omega (Jv, Jw) = \omega (v, w).
\end{equation}
\end{enumerate}
\end{definition}
Given this, one can naturally define a Riemannian metric $g \in \vee^2 \Gamma(TM)$ by:
\begin{equation}
g(v,w):= \omega(v, Jw).
\label{def:qm}
\end{equation}
In fact, a K\"ahler manifold can also be defined as a triple $(M, g, J)$ where $g$ is a Riemannian metric and $J$ is an invariant almost complex structure\footnote{%
Associated to the Riemannian metric, there is a unique torsion-free metric compatible affine connection $\nabla$ (the so-called Levi-Civita connection). An almost complex structure $J$ is said to be \emph{invariant} if $\nabla J = 0$\cite{kobayashi1969}.
}.
Equation \eqref{def:qm} is then perceived as the definition of the symplectic form. 

The important fact for us is that \emph{the quantum space of states is both a symplectic manifold and a Riemannian manifold}. It has thus a very rich geometry which can be used to provide an alternative description of the full Jordan-Lie algebra of quantum observables, with no reference to operators on the Hilbert space. This is achieved in two steps, as we explain in the subsequent subsections. Here, we follow closely \cite{ashtekar1997}.

\subsection{The symplectic-Lie structure of the quantum}
\label{quantum symplectic-Lie}

Let $\sh$ denote the collection of unit vectors of the Hilbert space $\hh$, and consider the pair of arrows
\begin{equation}
\begin{tikzcd}[column sep=large]
\hh \ar[r, hookleftarrow, "i"]& \sh \ar[r, twoheadrightarrow,"\tau"] & \PP\hh, 
\end{tikzcd}
\label{map:ph}
\end{equation}
where the left arrow is simply the injection saying that $\sh$ is a submanifold of $\hh$, and the right arrow is the projection describing the unit sphere as a $U(1)$-fibre bundle over the projective Hilbert space (in other words, it describes $\ph$ as a quotient: $\ph \simeq \sh/ U(1)$).

Consider now the map $\:\:\widehat \:\:: \bbr(\hh) \longrightarrow \cci(\ph, \RR)$ that, to a given self-adjoint operator $F$, associates the real-valued function defined by 
$$\widehat F(p) :=\langle \phi, F\phi\rangle, \text{ where } \phi \in \tau^{-1}(p),$$
and let us denote by $\cci(\ph, \RR)_\kk$ the image of this map. 

$\:\:\widehat\:\:\:$ is obviously an injection of real vector spaces. There is however no hope for this map to be a bijection, as may readily be seen by considering finite-dimensional Hilbert spaces: in this case, $\bbr(\hh)$ is finite-dimensional whereas $\cci(\ph, \RR)$ is infinite-dimensional. Through the map $\:\:\widehat\:\:\:,$ one can therefore think of self-adjoint operators as being some `very particular' functions on the projective Hilbert space. The difficult question is to specify what `very particular' means---\emph{i.e.} to characterize $\cci(\ph, \RR)_\kk$ inside $\cci(\ph, \RR)$. 

A first step in this direction is to use the symplectic structure of the quantum space of states. Let us quickly recall how it is defined. The easiest way is to start by noticing that the Hilbert space $\hh$ is itself a symplectic manifold. To see this, it is best to change perspectives and consider $\hh$ from the point of view of real numbers rather than complex numbers. First, one views $\hh$ as real vector space equipped with a complex structure $J$. This simply means that the multiplication of a vector by a complex number is now considered as the result of two operations---multiplication by real numbers and action of the linear operator $J$: for $z \in \CC$ and $\phi \in \hh$, we have $z \phi = \text{Re}(z) \phi + \text{Im}(z)J \phi$. Second, one also decomposes the Hermitian product of two vectors into its real and imaginary parts, and uses the natural isomorphism $T\hh \simeq \hh \times \hh$\footnote{%
Given $(\phi, \psi) \in \hh \times \hh$, define $V_\phi \in T_\psi\hh$ by 
\begin{equation*}
\forall f \in \cci(\hh,\RR), V_\phi[f](\psi)= \frac{d}{ds}f(\psi + s \phi)\Big|_{s=0}.
\end{equation*}
\vspace*{-0.8em}
}, 
to define the tensor $\Omega \in \Gamma(T^*\hh \otimes T^*\hh)$ by
\begin{equation}
\Omega(V_\phi, V_\psi):= 4\text{Im}(\langle \phi, \psi \rangle).
\label{def:hs}
\end{equation}
The skew-symmetry of the Hermitian product entails the anti-symmetry of $\Omega$, which is hence a 2-form. The fact that the Hermitian product is positive-definite and non-degenerate implies $\Omega$ is both closed and non-degenerate. Therefore, $\Omega$ is a symplectic structure on $\hh$. Given this, the symplectic form on the projective Hilbert space is the unique non-degenerate and closed 2-form $\omega \in \Omega^2(\ph)$ such that $\tau^*\omega = \iota^* \Omega$ (the pull-back of $\omega$ to the unit sphere coincides with the restriction to $\sh$ of the symplectic form on $\hh$)\footnote{%
Of course, one needs to be sure that such a 2-form does exist. A cleaner way of defining the symplectic form is by means of the so-called \emph{Marsden-Weinstein symplectic reduction} \cite{marsden1974}. Therein, one considers the natural action of $U(1)$ on $\hh$. This is a strongly Hamiltonian action and the momentum map $\mu: \hh \longrightarrow u(1)^* \simeq \RR$ is given by $\mu(\phi)=\langle \phi, \phi \rangle$. Then, $\ph \simeq \mu^{-1}(1)/U(1)$ and the general theory insures this is a symplectic manifold. For the details, I refer the reader to \cite{landsman1998}.
}.
The induced Poisson bracket on $\cci(\ph, \RR)$ will be denoted by $\{\cdot, \cdot\}_{\ph}$.

The question now is whether the symplectic structure plays in quantum kinematics exactly the same role as it does in classical kinematics---namely, whether it allows to define both the Lie product on the algebra of quantum observables, and the generator of state transformations associated to any given observable. As it turns out, the answer is positive. Indeed, it can be shown \cite{ashtekar1997, landsman1998} that the Hamiltonian vector field on $\ph$ associated to the self-adjoint operator $F$ (regarded through the map $\:\:\widehat\:\:\:$ as a function on $\ph$) coincides with the projection of the vector field $V_F$ (which is defined on $\hh$ (cf. \eqref{map:qot}) but is in fact tangent to $\sh$ since it generates unitary transformations). In other words, we have
\begin{equation}
\forall F \in \bbr(\hh), \: v_{\widehat F}=\tau_*V_F \in \Gamma(T\ph).
\label{eq:qvf}
\end{equation}
Moreover, we also have
\begin{equation}
\{\widehat F, \widehat K\}_{\ph} = \frac{i}{2}\widehat{[ F,  K]}
\label{eq:pbelp}
\end{equation}
which means that the map $\:\:\widehat\:\:\:$ is an injection of Lie algebras:
\begin{eqnarray}
\begin{tikzcd}[]
(\bbr(\hh), \frac{i}{2}[\cdot, \cdot]) \ar[r, hookrightarrow, "\widehat\:"] &(\cci(\ph, \RR), \{\cdot, \cdot\}_{\ph}).
\end{tikzcd}
\label{inj:bf}
\end{eqnarray}
Hence, the commutator of bounded self-adjoint operators may be seen as the restriction to $\cci(\ph, \RR)_\kk$  of the Poisson bracket induced by the symplectic structure on $\ph$. Together, equations \eqref{eq:qvf} and \eqref{eq:pbelp} show that, \emph{as far as the Lie structure and the transformational role of quantum observables is concerned, we might as well forget self-adjoint operators and reason in terms of expectation-value functions and the intrinsic symplectic geometry of the quantum space of states}.

This should be felt as an impressive merger of the two kinematics. Any space of states, be it classical or quantum, is a symplectic manifold and the symplectic structure plays exactly the same role in both cases: it induces the Lie structure on the algebra of observables and governs their transformational role. Or, to put it differently, if one decides to restrict attention and focus only on the transformational role of observables, then there is no difference whatsoever between classical and quantum kinematics. In particular, any statement of classical mechanics which only involves observables-as-transformations goes unchanged when passing to quantum mechanics\footnote{%
Two examples of this are the canonical commutation relations $\{p,q\}=1$ (which state that linear momentum is the generator of space translations) and Hamilton's equations of motion $\frac{d}{dt}=\{H, \cdot\}$ (which state that the Hamiltonian is the generator of time evolution).
}.

\subsection{The Riemannian-Jordan structure of the Quantum}
\label{quantum Riemannian-Jordan}

Despite the injection \eqref{inj:bf}, the full \emph{Jordan}-Lie algebra $(\bbr(\hh),  \frac{i}{2}[\cdot, \cdot],  \frac{1}{2}[\cdot, \cdot]_+)$ cannot be seen as a subalgebra of the Poisson algebra $(\cci(\ph, \RR), \{\cdot, \cdot\}_{\ph}, \cdot)$. The obstruction, of course, lies on the Jordan structure: although quantum observables may be represented as functions on the space of states, the associative point-wise multiplication of functions cannot yield the non-associative Jordan product of quantum observables.

Another simple way of understanding the obstruction is to reflect on the representation of the \emph{square} of an observable\footnote{%
The reader will recognise here the question raised by Heisenberg in his 1925 seminal paper that definitely launched the development of quantum mechanics\cite{heisenberg1925}.  
}. 
Suppose that, for any given observable $\texttt{f}$ represented by the abstract element $f$, it is known how to construct the abstract element $f^2$ representing the observable $\texttt{f}^2$ (operationally defined as the observable associated with squaring the numerical results of all measurements of $\texttt{f}$). Then, this operation of taking squares allows us to define a Jordan product $\bullet$ on the algebra of observables by 
$$f\bullet k:=\frac{1}{4}((f+k)^2 -(f-k)^2).$$
Thus, in general, given an observable $\texttt{f}$ represented by the abstract element $f$, its square $\texttt{f}^2$ should be represented by the element $f \bullet f$, where $\bullet$ is the Jordan product. In classical kinematics, if $\texttt{f}$ is represented by a certain real-valued function $f$, then the observable $\texttt{k}:=\texttt{f}^2$ is represented simply by the square function $f^2$. But, as we know, this is not the answer of quantum kinematics: the observable $\texttt{f}$ is represented by the self-adjoint operator $F$ and the observable $\texttt{k}$ by the self-adjoint operator $F^2$. Thus, in terms of the functions over the space of states, we have $k = \widehat{F^2} \neq (\widehat F)^2 = f^2$. This shows that indeed point-wise multiplication is not the right structure for the quantum case. Instead, one should try to define a Jordan product $\bullet$ satisfying $\widehat{F^2} = \widehat F \bullet \widehat F$. 

With little surprise, this is achieved by using the additional geometric structure present on the quantum space of states that we have ignored so far: the Riemannian metric $g$. This structure is defined in very similar fashion to the construction of the symplectic form. Now, one considers the real part of the Hermitian product to define the tensor $G \in \Gamma(T^*\hh \otimes T^*\hh)$ by: 
\begin{equation}
G(V_\phi, V_\psi):= 4\text{Re}(\langle \phi, \psi \rangle).
\label{eq:hilbert_riemann}
\end{equation}
This time, the skew-symmetry, positive-definitiveness and non-degeneracy of the Hermitian product respectively imply the symmetry, positive-definitiveness and non-degeneracy of $G$, which is hence a Riemannian metric on $\hh$. 

At this point, we may use again diagram \eqref{map:ph} to induce a Riemannian metric on the space of states. In the symplectic case, we regarded the isomorphism $\sh/U(1) \simeq \ph$ as the second stage of the Marsden-Weinstein symplectic reduction and this sufficed to insure $\ph$ was also symplectic. Instead, we now adopt towards this isomorphism a different perspective, called by Ashtekar and Schilling the ``Killing reduction''\cite{ashtekar1997}. It is the following: first, the restriction $i^*G$ of the metric $G$ to the unit sphere is again a metric and $\sh$ becomes then a Riemannian manifold in its own right. Second, one regards the action of $U(1)$ on $\hh$ as the one-parameter group of transformations generated by the vector field $V_{Id} \in \Gamma(T\hh)$ associated to the identity self-adjoint operator. Since these transformations preserve the Hermitian product, they also preserve the metric $G$. Thus $V_{Id}$ is a Killing vector. Moreover, this vector field is tangent to $\sh$ and is hence also a Killing vector for $i^*G$. In this way, the isomorphism $\sh/U(1) \simeq \ph$ describes the projective Hilbert space as the space of all trajectories of the Killing vector field $V_{Id}$. By a result of Geroch\cite{geroch1971}, we know that the resulting manifold is also Riemannian. The Riemannian metric on $\ph$ is called the \emph{Fubini-Study} metric and will be denoted by $g$\footnote{%
As a side remark, notice that, in the same way that the Riemannian and symplectic structures of the quantum space of states arise then from the real and imaginary parts of the Hermitian product of $\hh$ respectively, at the algebraic level the quantum Jordan and Lie products $\bullet : \frac{1}{2}[\cdot, \cdot]_+$ and  $\star : \frac{i}{2}[\cdot, \cdot]$ may also be seen as the real and imaginary parts of the composition of operators:$$\text{for } A, B \in \bbr(\hh), \: A \circ B = A \bullet B - i A\star B.$$ 
\vspace*{-1.7em} 
}.  

Given this metric, and in very similar fashion to the definition of the Poisson bracket in terms of the symplectic structure (cf. eq \eqref{def:pb}), one can define the following product on $\cci(\ph, \RR)$: 
\begin{equation}
\label{def:rb}
\forall f, k \in \cci(\PP \hh, \RR), \:\: f \bullet k := g(v_f, v_k) + f\cdot k,
\end{equation}
where, to the point-wise multiplication of functions $f\cdot k$, the metric adds a ``Riemannian bracket'' $(f, k) :=g(v_f, v_k)$. The result is a commutative and non-associative product. Thus, the presence of the Riemannian structure allows to \emph{deform} the usual commutative and associative algebra of functions into a commutative but non-\-asso\-ciative algebra. Yet, this does not turn $(\cci(\ph, \RR), \bullet, \{\cdot, \cdot\}_{\ph})$ into a Jordan-Lie algebra, for $\bullet$ and $\{\cdot, \cdot\}_{\ph}$ do not satisfy the associator rule in general. 

Remarkably, however, one has the identity\cite{ashtekar1997}:
\begin{equation}
\label{eq:qrj}
\forall F, K \in \bbr(\hh), \widehat F \bullet \widehat K = \frac{1}{2}\widehat{[F,K]_+}.
\end{equation}
which has many important implications. First, it implies $\widehat F \bullet \widehat F = \widehat{F^2}$, as we wanted. Second, it shows that $(\cci(\ph, \RR)_\kk, \bullet)$ is a subalgebra of $(\cci(\ph, \RR), \bullet)$ and, more importantly, that \emph{when restricted to this subalgebra} the new product $\bullet$ becomes a Jordan product. In other words, we now have the isomorphism of non-associative Jordan-Lie algebras: 
\begin{eqnarray}
\boxed{
\big(\bb_\RR(\hh), \frac{1}{2}[\cdot, \cdot]_+, \frac{i}{2}[\cdot, \cdot]\big) \simeq \big(\cci(\ph, \RR)_\kk, \bullet, \{\cdot, \cdot\}_{\PP\hh}\big).
}
\label{iso:operators_functions}
\end{eqnarray}

In addition to its role in the definition of the Jordan product for quantum observables, the presence of a metric in the quantum space of states provides a very simple geometric interpretation of two other crucial aspects of quantum kinematics: its transition probability structure and the indeterminacy in the numerical value of observables. First, given a state $p$, the probability that a measurement of the observable $\widehat F$ will yield the result $\lambda$ is given by 
\begin{equation}
\text{Pr}(p, \widehat F = \lambda) = \cos^2\big(d_g(p, \Sigma_\lambda)\big)
\end{equation}
where $\Sigma_\lambda$ is the subset of states having $\lambda$ as definite value of the observable $\widehat F$, and $d_g(p, \Sigma_\lambda)$ is the minimal distance between the state $p$ and the subset $\Sigma_\lambda$\footnote{%
Recall that the distance between two points $p$ and $p'$ of a Riemannian manifold with metric $g$ is given by: $$d_g(p,p') := \text{inf}\:\big\{\int\limits_\Gamma \sqrt{g(v_\Gamma(t), v_\Gamma(t))}dt \: \big| \: \Gamma \in \text{Path}(p,p')\big \}.$$ 
}.
In other words, the quantum transition probabilities appear here to be simply a measure of the distance in the quantum space of states. Second, from the combination of \eqref{def:rb} and \eqref{eq:qrj}, we get
\begin{equation}
\label{eq:gu}
\Delta F = \widehat F \bullet \widehat F - \widehat F \cdot \widehat F = g(v_F, v_F), 
\end{equation}
which shows that the uncertainty of a quantum observable is nothing but the norm of the Hamiltonian vector field associated to it. 

From the conceptual perspective that is ours, this last result is particularly enlightening. Indeed, in terms of the two-fold role of observables, this can also be expressed as: given a state $\rho$ and an observable $F$, \emph{the uncertainty $\Delta F(\rho)$ in the numerical value of the observable $F$ is precisely a measure of how much the state $\rho$ is changed by the transformations generated by the observable}. In particular, we recover as a special case the relation, noted at the end of \autoref{standardQ} between definite-valuedness and invariance. Thus, it brings to the fore the existence in quantum kinematics of an interdependence between the numerical and transformational role of observables which is absent in classical kinematics.

\subsection{The geometric characterization of quantum observables}
\label{geometric quantum observables}

Although we have now reached a completely geometric definition of both the Jordan and Lie structures governing the algebra of quantum observables, the reference to operators on Hilbert spaces has not yet been eliminated altogether: the elements of the algebra $\cci(\ph, \RR)_\kk$ are still defined as expectation-value functions associated to bounded self-adjoint operators. Thus, the last stone in the full geometric description of quantum observables is to furnish a criterion allowing to know when a function $f \in \cci(\ph, \RR)$ is of this form. 

Many different characterizations exist\cite{landsman1994, shultz1982, alfsen1976}, but the simplest one---and the most relevant one from the point of view of the two-fold role of observables---was found first by Schilling\cite{schilling1996} (and shortly later rediscovered by Cirelli, Gatti and Manià\cite{cirelli1999}). 
  
\begin{theorem}
\label{thm:obs}
Let $f$ be a smooth real-valued function over the projective Hilbert space $\ph$. The following two conditions are equivalent:
\begin{enumerate}[leftmargin=1cm, label=\roman*)]
\item there exists a bounded self-adjoint operator $F \in \bbr(\hh)$ such that, for any $\phi \in \sh$, $f([\phi]) = \langle \phi, F\phi\rangle$,
\item the Hamiltonian vector field associated to the function $f$ is also a Killing vector field. 
\end{enumerate}
\end{theorem}

In one direction, the equivalence is obvious. In the other, the proof is essentially a combination of Wigner's theorem \cite{wigner1959, bargmann1954} (which forces the one-parameter group of transformations generated by the Killing field to be a group of unitary transformations) and Stone's theorem (which forces the unitary transformations to be generated by a self-adjoint operator). The delicate part of the proof is to show that indeed it is a \emph{bounded} self-adjoint-operator. 

In this way, one reaches a mathematical definition of observables that applies equally well to classical and quantum kinematics:
\medskip
\begin{center}
\noindent\fbox{
\parbox{0.9\linewidth}{
\noindent	\textbf{Observables}: an observable of a physical system is a smooth real-valued \emph{function} on the space of states $S$ to which an \emph{infinitesimal state transformation} can be associated. That is, it is a function whose associated vector field preserves \emph{all} the geometric structures present in the space of states.
}}
\end{center}
\phantomsection
\label{def:o}
\medskip

In the classical case, there is only the symplectic structure to preserve. Hence, any function $f$ does the job, since its Hamiltonian vector field $v_f$ automatically verifies $\ll_{v_f}\omega = 0$. But in the quantum case, there is also the metric to preserve. Accordingly, only those functions for which the symplectic gradient is also a Killing vector field will qualify as observables. Theorem \ref{thm:obs} guarantees that these functions exactly coincide with the functions $\widehat F$ that are real expectation-value maps of bounded self-adjoint operators $F$. Moreover, it is important to notice that this last point only applies to the \emph{projective} Hilbert space. Were one to insist on working at the level of $\hh$, this geometric characterization of observables would fail, for there are too many functions preserving both the symplectic and Riemannian structures which do not arise as expectation-value maps of operators\cite{ashtekar1997}.

The conceptual relevance of this definition should not be missed, as it enlightens the essential importance of the two-fold role of observables in kinematics. For \emph{it is precisely this two-fold role, numerical and transformational, that serves as a definition of what an observable is}. The standard definition of classical observables only involved their numerical role---they were defined as functions on the space of states---and did not apply to quantum kinematics. Conversely, the standard definition of quantum observables only involved their transformational role---they were defined as operators acting on states---and did not apply to classical kinematics. A posteriori, it is therefore most natural that the general mathematical definition of an observable, classical or quantum, should explicitly mention both the functions and the transformations. 

\subsection{Classical vs Quantum in the geometric formulation}
\label{c vs q geometric}

\autoref{tbl:quantum classical geometric} (\autopageref{tbl:quantum classical geometric}) below summarizes the conceptual understanding of classical and quantum kinematics that emerges from the analysis of their geometric formulations.

\begin{table}[!ht]
\begin{center}
\renewcommand{\arraystretch}{1.8}
\small\linespread{1.2}\begin{tabular}{ |c||c|c|}

 \hline
 & \parbox{14em}{\centering \textbf{Classical Kinematics}} 
 & \parbox{14em}{\centering \textbf{Quantum Kinematics}} 
 
 \\
\hline
\hline

\parbox{6em}{\centering \textbf{States}} 
& \parbox{14em}{\centering ~\\ points $p$ of a symplectic manifold \\ $(M,\omega)$\\ ~  }
& \parbox{14em}{\centering ~\\points $p$ of a K\"ahler manifold \\ $(M, \omega, g, J)$\\ ~}

 \\
 \hline

\parbox{6em}{\centering \textbf{Observables}}
 & \parbox{15em}{\centering ~\\ $\cci(M, \RR)_\kk$\\ Smooth real-valued functions whose transformations preserve the geometric structures\\ ~}
 & \parbox{15em}{\centering ~\\ $\cci(M, \RR)_\kk$\\ Smooth real-valued functions whose transformations preserve the geometric structures\\ ~}

 \\
 \hline 
\parbox{6.5em}{\centering \textbf{~\\Geometric structures of states}\\~}
 & \parbox{15em}{
 \textcolor{blue}{$\spadesuit$} a symplectic 2-form $\omega$ \\[0.5em] $\:$}
 & \parbox{15em}{
 \textcolor{blue}{$\spadesuit$}  a symplectic 2-form $\omega$ \\[0.5em] 
 \textcolor{red}{$\clubsuit$} \textbf{a Riemannian metric} $g$}

 \\
 \hline 

 \parbox{6.5em}{\centering \textbf{Algebraic structures of observables}}
 & \parbox{15em}{  
 \textcolor{blue}{~\\$\spadesuit$}  Anti-commutative Lie product\\[0.2em]  
\hspace*{3.7em}$\{f,k\} = \omega(V_f, V_k)$ \\[0.2em] 
\hspace*{2.6em}(induced by symplectic)\\[1em] 
\textcolor{red}{$\clubsuit$} Jordan product \\[0.2em] 
Commutative \textbf{and associative} \\[0.2em]  
\hspace*{4.5em}$f \bullet k = f\cdot k$\\[0.2em] ~\\ ~}
 
 & \parbox{15em}{ 
 \textcolor{blue}{~\\ $\spadesuit$}  Anti-commutative Lie product\\[0.2em]  
\hspace*{3.7em}$\{f,k\} = \omega(V_f, V_k)$ \\[0.2em] 
\hspace*{2.6em}(induced by symplectic)\\[1em] 
\textcolor{red}{$\clubsuit$} Jordan product \\[0.2em] 
Commutative \textbf{but non-associative} \\[0.2em]  
\hspace*{3em}$ f\bullet k = f\cdot k + g(V_f, V_k)$\\[0.2em] 
\hspace*{2.6em}(induced by Riemannian)\\ ~}
 
 \\
 \hline 


\parbox{6em}{\centering \textbf{Two-fold \\role of observables}}
 & \parbox{15em}{~\\
 \textcolor{blue}{$\spadesuit$}  Transformational role captured by\\[0.2em]  
\hspace*{5em}Lie product \\[0.6em] 
\textcolor{red}{$\clubsuit$} Quantitative role captured by\\[0.2em]
\hspace*{4em}Jordan product\\~}

 & \parbox{15em}{~\\
 \textcolor{blue}{$\spadesuit$}  Transformational role captured by\\[0.2em] 
\hspace*{5em}Lie product \\[0.6em]
\textcolor{red}{$\clubsuit$} Quantitative role captured by \\[0.2em]  
\hspace*{4em}Jordan product\\~}
 
 \\
 \hline 
 
\parbox{6em}{\centering \textbf{Link between the two roles}}
 & \parbox{15em}{\centering ~\\ Numerical role \textbf{independent} of transformations:\\[0.2em] 
 $\Delta f= 0$\\~}
 & \parbox{15em}{\centering ~\\ Numerical role \textbf{dependent} on transformations:\\[0.2em]
$\Delta f = g(v_f, v_f)$\\~}
 \\
 \hline
\end{tabular}
\normalsize
\end{center}
\vspace*{-1em}
\caption{Comparison of the main mathematical structures present in classical and quantum kinematics (geometric formulation).}
\label{tbl:quantum classical geometric}
\end{table}

The comparison is striking: in contrast to the staggering difference between the two theories conveyed by the standard formalisms, the geometric point of view highlights the deep common ground shared by the classical and the quantum. It allows, by the same token, to pinpoint more precisely the place where they differ. Following Schilling, it is indeed tempting to say that 
\begin{quote}
the fundamental distinction between the classical and quantum formalisms is the presence, in quantum mechanics, of a Riemannian metric. While the symplectic structure serves exactly the same role as that of classical mechanics, the metric describes those features of quantum mechanics which do not have classical analogues.\cite[p. 48]{schilling1996}
\end{quote} 
Both the non-associativity of the Jordan product and the indeterminacy of the values for quantum observables explicitly involve the metric. This view---that the quantum world has one additional geometric structure, \emph{with no analogue in the classical},  and that, in a loose sense, to quantize is to add a Riemannian metric to the space of states---is found in the vast majority of works which played an important role in developing the ``geometrization or delinearization program" of quantum mechanics. 

Nonetheless, this is \emph{not} the impression conveyed by the comparative table. It is not so much that the classical analogue of the Riemannian structure is \emph{missing} but rather, one would be tempted to say, that the classical analogue is \emph{trivial}. Indeed, one gets the impression that the correct classical `Riemannian metric'' is $g=0$, for setting $g$ to vanish in the quantum formulas yields the classical ones. Of course, $g=0$ is not an actual metric, but this does suggest there may be yet another manner of formulating the two kinematical arenas, a manner in which they both exhibit the same two kinds of geometric structures, and it just so happens that one of these structures is trivial---and hence unnoticed---in classical kinematics.


\section{Landsman's axiomatization of quantum mechanics}
\label{landsman}

In the geometric approach to mechanics, the goal of a unifying programme is to find a notion of space which meets the following three requirements:
\begin{enumerate}[label=\arabic*)]
\item \emph{Unification of states}: both the classical and quantum space of states fall under the same notion of space.
\item \emph{Unification of observables:} there is a unique definition of the algebra of observables, which, when restricted to the classical case, yields a Poisson algebra, and when restricted to the quantum case yields a non-associative Jordan-Lie algebra.
\item \emph{Characterization of the quantum}: there is a physically meaningful characterization of when a space of this sort is a quantum space of states. 
\end{enumerate}
As we have just seen, the formulation of quantum mechanics in terms of K\"ahler manifolds achieves the second requirement but it fails to meet the first one (and thus the last one).

At the end of the last century, Nicolaas P. Landsman developed an alternative approach which succeeds in meeting the three demands\cite{landsman1997, landsman1998, landsman1998c}. One can consider that the starting point of his approach is to extend the geometric formulation of mechanics to the case where there exist superselection rules. In this situation, the quantum space of states is no longer described by a single projective Hilbert space $\ph$ but, instead, by a disjoint union of many: $\pp^Q= \sqcup_\alpha \ph_\alpha$. The classical analogue of this is to consider general Poisson manifolds instead of focusing only on symplectic manifolds\footnote{%
A \emph{Poisson manifold} is a manifold $P$ for which the algebra of smooth functions $\cci(P, \RR)$ is a Poisson algebra. An important theorem in Poisson geometry states that any such manifold can always be written as a disjoint union of symplectic manifolds---the so-called \emph{symplectic leaves} of the Poisson manifold\cite[Theorem I.2.4.7, p. 71]{landsman1998}.  
}. 

However, by performing such an extension, the Riemannian metric does no longer suffice to define all the transition probabilities on the quantum space of states. The problem arises when considering two inequivalent states $p$ and $p'$ (that is, two states belonging to different superselection sectors $\ph_\alpha$ and $\ph_{\alpha'}$): the geometric formula $\text{Pr}(p, p') = \cos^2(d_g(p, p'))$ cannot be applied since there is no notion of distance between points of different sectors. In the light of this, the natural strategy is to reverse the priority between the metric and the transition probabilities: instead of considering the metric $g$ as a primitive notion and the transition probabilities as a derived notion, take the transition probabilities as a fundamental structure of the quantum space of states.

\subsection{Poisson spaces with transition probability}

The relevant notion of space coined by Landsman is that of a  \emph{Poisson space with a transition probability}. In preparation of the definition of this notion, we first need to introduce some terminology:

\begin{definition}
\label{def:ps}
A \emph{Poisson space} is a Hausdorff topological space $\pp$ together with a collection $S_\alpha$ of symplectic manifolds, as well as continuous injections $\iota_\alpha: S_\alpha \hookrightarrow \pp$, such that\index{Hausdorff spaces}
\vspace*{-0.4em}
$$\pp = \bigsqcup\limits_{\alpha} \iota_\alpha(S_\alpha).\vspace*{-0.7em}$$
The subsets $\iota_\alpha(S_\alpha) \subset \pp$ are called the \emph{symplectic leaves} of $\pp$\footnote{%
This notion was introduced for the first time by Landsman in \cite[p. 38]{landsman1997}. His definition differs slightly from the one given here, for it also includes a linear subspace $\uur(\pp) \subset \cc^\infty_L(\pp, \RR)$ which separates points and is closed under the Poisson bracket: $\{f, g\}_{\raisebox{-2pt}{$\scriptstyle \pp$}}(\iota_\alpha(q)) := \{\iota^*_\alpha f, \iota^*_\alpha g\}_{\raisebox{-2pt}{$\scriptstyle S_\alpha$}}(q)$, where $q \in S_\alpha$. I nonetheless find the inclusion of this subspace somewhat unnatural at this point. This subspace $\uur(\pp)$ will only become important when defining the key notion of a Poisson space with transition probability.
}. 
\end{definition}

\begin{definition}
A symmetric \emph{transition probability space} is a set $\pp$ equipped with a function $\text{Pr}: \pp \times \pp \longrightarrow [0,1]$ such that for all $\rho, \sigma \in \pp$\index{transition probability|textbf}\index{transition probability! space|textbf} 
\begin{enumerate}[label=\roman*)]
\item $\text{Pr}(\rho, \sigma) = 1 \:\Longleftrightarrow \rho = \sigma$,
\item $\text{Pr}(\rho, \sigma) = \mathrm{Pr}(\sigma, \rho)$ (i.e. $\text{Pr}$ is symmetric).
\end{enumerate}
The function $\text{Pr}$ is called a \emph{transition probability}\footnote{%
This concept was introduced for the first time in 1937 by von Neumann in a series of lectures delivered at the Pennsylvania State College. The manuscript was only published posthumously in 1981\cite{neumann1981}. 
}. 
\end{definition}

Now, recall the geometric characterization of quantum observables achieved in \autoref{thm:obs}: the quantum space of states was seen to be endowed with two geometric structures---a symplectic form $\omega$ and a Riemannian metric $g$. Associated to the symplectic structure $\omega$ was the set of functions $\cci(\ph, \RR)_\omega$ preserving it. Similarly, to the Riemannian metric $g$ one associated the set $\cci(\ph, \RR)_g$. Then, the algebra of observables was simply found to be the intersection:
\vspace*{-0.6em}
$$\cci(\ph, \RR)_\kk:= \cci(\ph, \RR)_\omega \cap \cci(\ph, \RR)_g.\vspace*{-0.4em}$$
This idea may be immediately transposed for those spaces $\pp$ which are equipped with the two structures just defined: one considers the function space $\cc_{Prob}(\pp, \RR)$ intrinsically related to a transition probability space and the function space $\cci_{Pois}(\pp, \RR)$ intrinsically associated to a Poisson space\footnote{%
These function spaces are defined as follows. First, $\cci_{Pois}(\pp, \RR)$ is the set of all $f \in \cc(\pp, \RR)$ such that their restrictions to any $S_\alpha$ are smooth: $\iota^*_\alpha f \in \cci(S_\alpha, \RR)$. On the other hand, the definition of $\cc_{Prob}(\pp, \RR)$ is more involved. One considers first the functions $\mathrm{Pr}_\rho : \pp \rightarrow \RR$ such that $\mathrm{Pr}_\rho (\sigma) := \mathrm{Pr} (\rho, \sigma)$, and defines $\cc_{Prob}^{00}(\pp)$ as the real vector space generated by these functions. Then $\cc_{Prob}(\pp,\RR):=  \overline{\cc_{Prob}^{00}(\pp)}^{**}$. See \cite[pp. 76--84]{landsman1998} for more details.
} 
in order to define
\vspace*{-0.2em}
\begin{eqnarray}
\cc(\pp, \RR)_\kk :=  \cci_{Pois}(\pp, \RR) \cap \cc_{Prob}(\pp, \RR).
\label{Landsman observables}
\end{eqnarray}

We are now ready to introduce the central definition:
\begin{definition}
\label{def:upswtp}
A \emph{Poisson space with a transition probability} is a set that is both a transition probability space and a Poisson space and for which $\cc(\pp, \RR)_\kk$, as defined in \eqref{Landsman observables}, satisfies:
\begin{enumerate}[label=\roman*)]
	\item completeness: $\cc(\pp, \RR)_\kk$ separates points,
	\item closedness: $\cc(\pp, \RR)_\kk$ is closed under the Poisson bracket,
	\item unitarity: the Hamiltonian flow defined by each element of $\cc(\pp, \RR)_\kk$ preserves the transition probabilities.
\end{enumerate}
\label{def:Poisson space with probability}
\end{definition}

From \autoref{geometric}, it is clear that a projective Hilbert space, equipped with its natural symplectic form and the transition probability function $\text{Pr}(p, p')=\cos^2(d_g(p,p'))$ induced by the Fubini-Study metric $g$, satisfies all three axioms and qualifies hence as a Poisson space with transition probability. On the other hand, one can always consider any Poisson manifold---and in particular any symplectic manifold $S$--- as a Poisson space with transition probability, where the transition probability function is trivial: $\text{Pr}(p,p') = \delta_{p,p'}$. In this case, we have $\cc_{Prob}(S, \RR) = \cc(S, \RR)$, $\cc(S, \RR)_\kk = \cci(S, \RR)$ and the three axioms are trivially met. 

As it has been the case for all other structures that we have met in the description of kinematics, the fundamental notion introduced by Landsman to achieve the geometric unification of classical and quantum kinematics is a space endowed with \emph{two} structures. To show that these are the geometric counterparts of the two algebraic structures present on the algebra of observables, what remains to be seen is how to construct a Jordan product on $\cc(\pp, \RR)_\kk$ starting from a transition probability function $\text{Pr}$.     

This is achieved by noticing that, for transition probability spaces, one can develop a spectral theory, much in the like of the spectral decomposition of self-adjoint operators on a Hilbert space. Given a transition probability space $\pp$, define a \emph{basis} $\bb$ as an orthogonal family of points in $\pp$ such that $$\sum\limits_{\rho \in \bb} p_\rho=1,$$ 
where $p_\rho$ is the function on $\pp$ defined by $p_\rho(\sigma):=\text{Pr}(\rho, \sigma)$\footnote{%
Given a transition probability space $(\pp, \mathrm{Pr})$, two subsets $\ss_1$ and $\ss_2$ are said to be \emph{orthogonal} if, for any $p \in \ss_1$ and any $p' \in \ss_2$, $\mathrm{Pr}(p, p') = 0$. A subset $\ss \subset \pp$ is said to be a \emph{component} if $\ss$ and $\pp \setminus \ss$ are orthogonal. Finally, a \emph{sector} is a component which does not have any non-trivial components.
}. 
It can be shown \cite[Proposition I.2.7.4]{landsman1998} that all the bases of $\pp$ have the same cardinality and hence allow to define a notion of `dimension' for a transition probability space. Now, given an orthoclosed subset $\ss \subset \pp$\footnote{%
Given a subset $\ss \subset \pp$, the orthoplement $\ss^\bot$ is defined by 
\vspace*{-0.3em}
$$\ss^\bot :=\big\{ p \in \pp\:\big|\: \forall s \in \ss, \text{ Pr}(p, s) =0\big\}.\vspace*{-0.5em}$$
In turn, a subset is called \emph{orthoclosed} whenever $\ss^{\bot \bot} = \ss$.
} 
and a basis $\bb$ thereof, define the function $p_\ss:=\sum_{\rho \in \bb}p_\rho$. This function turns out to be independent of the choice of the basis $\bb$\footnote{
To be more precise, this holds only for \emph{well-behaved} transition probability spaces. A transition probability space is said to be well-behaved if every orthoclosed subset $\ss \subset \pp$ has the property that any maximal orthogonal subset of $\ss$ is a basis of it. See \cite[Definition I.2.7.5 and Proposition I.2.7.6]{landsman1998}.
}. 
With this in hand, we can now define the spectral theory:
\begin{definition}
Consider a well-behaved transition probability space $(\pp, \text{Pr})$ and a function $A \in \ell^\infty(\pp, \RR)$. Then a \emph{spectral resolution} of $A$ is an expansion
$$A = \sum\limits_{j} \lambda_j p_{\ss_j}$$
where $\{\ss_j\}$ is an orthogonal family of orthoclosed subsets of $\pp$ such that $\sum p_{\ss_j}=1$.
\end{definition}
The crucial point which confers to the spectral resolution its power is the fact that, for both Poisson manifolds (equipped with the trivial transition probabilities) and spaces of the form $\pp = \bigcup \ph_\alpha$, the spectral resolution is \emph{unique} and can thus be used to define the square of an observable by
\begin{eqnarray}
A^2:=\sum\limits_{j} (\lambda_j)^2 p_{\ss_j}.
\label{Jsq}
\end{eqnarray}
Finally, this allows to define the Jordan product by
\begin{eqnarray}A \bullet B := \frac{1}{4}\big((A+B)^2 - (A-B)^2\big).
\label{Jdef}
\end{eqnarray}

In sum, the notion of a Poisson space with transition probability succeeds in providing a common geometric language in which to describe both classical and quantum state spaces (requirement of unification of states), and from which one can construct, through a unified procedure, the algebra of classical or quantum observables (requirement of unification of observables). We now turn to the last and most important point: the characterization of the class of Poisson spaces with transition probability which describe quantum systems. 

\subsection{Characterization of quantum kinematics}

As it has been hinted at several times, from the point of view of the two-fold role of observables, a fundamental difference between classical and quantum kinematics seems to lie in the connection between the two roles: while in classical kinematics the two roles seem to be essentially independent from each other, in quantum kinematics the quantitative aspect of observables encodes information about the transformational one. Therefore, it is interesting to compare the behaviour of the Poisson structure and the transition probability structure. In order to do so, consider the following two equivalence relations defined on any Poisson space with transition probability:
\begin{definition}
\label{def:er}
Let $\pp$ be a Poisson space with transition probability. Then, two points $p, p' \in \pp$ are said to be:
\begin{enumerate}
\item \emph{transformationally equivalent}, denoted by $p \underset{T}{\sim} p'$, if they belong to the same symplectic leave,

\item \emph{numerically equivalent}, denoted by $p \underset{N}{\sim} p'$, if they belong to the same probability sector.
\end{enumerate}
\end{definition}

These two different equivalence relations may be seen as the two different notions of \emph{connectedness} of the space of states arising from the two fundamental geometric structures. `Transformational equivalence' is connectedness from the point of view of the transformational role of observables: two states $p$ and $p$ are transformationally equivalent if and only if there exists a curve $\gamma$ on $\pp$ generated by an element $F \in \cc(\pp, \RR)_\kk$ such that $p, p' \in \gamma$. In similar fashion, `numerical equivalence' is connectedness from the point of view of transition probabilities: two states are numerically equivalent if and only if there exists a collection of intermediate states $\chi_1, \ldots, \chi_n$ such that the chain of transitions $p \rightarrow \chi_1 \rightarrow \ldots \rightarrow \chi_n \rightarrow p'$ has a non-vanishing probability---\emph{i.e.}, such that $\mathrm{Pr}(\rho, \chi_1)\mathrm{Pr}(\chi_1, \chi_2) \ldots \mathrm{Pr}(\chi_{n-1},\chi_n)\mathrm{Pr}(\chi_n,\sigma) \neq 0$. 

Now, in classical kinematics, where one considers as space of states $\pp_{cl}$ a symplectic manifold with transition probabilities $\mathrm{Pr}(p, p')=\delta_{p, p'}$, the two notions of connectedness are at odds from each other: from the point of view of the Poisson structure, the space of states is completely connected (any two states are transformationally equivalent), whereas from the point of view of the transition probability structure the space of states is completely disconnected (no two different states are numerically equivalent). In other words, we have
\vspace*{-0.5em} 
$$\ast = (\pp_{cl} / \underset{T}{\sim})  \neq (\pp_{cl} / \underset{N}{\sim}) = \pp_{cl}.
\vspace*{-0.5em}$$ 
On the other hand, in quantum kinematics the compatibility between the two roles of observables is captured in the fact that these two a priori different equivalence relations coincide. Indeed, we have
$$(\pp_{qu} / \underset{T}{\sim})  = (\pp_{qu} / \underset{N}{\sim}).$$ 

One of the great achievements of Landsman's approach in terms of Poisson spaces with transition probabilities is to show with unmatched clarity that this compatibility between the two roles of observables is in fact one of the essential differences between classical and quantum kinematics. Indeed, given a Poisson space with a transition probability, he has provided the following axiomatic characterization of when such a space is a quantum space of states: 

\begin{theorem}
A non-trivial Poisson space with a transition probability $\pp$ is the pure state space of a non-commutative $C^*$-algebra if and only if:
\begin{enumerate}[label=QM\:\arabic*), leftmargin=5em]
\item %
\label{PoS} %
\textbf{Principle of superposition}: \\for any $p, p'\in \pp$ such that $p \underset{N}{\sim} p'$ and $p\neq p'$, we have $\{p, p'\}^{\bot\bot}\simeq S^2$.

\item \textbf{Compatibility of the two roles of observables}: \\the probability sectors and the symplectic leaves of $\pp$ coincide.\footnote{%
Here, a ``trivial Poisson space'' means a Poisson space whose Poisson bracket is identically zero. As stated, the theorem is only valid for \emph{finite}-dimensional $C^*$-algebras. In the infinite-dimensional case, two more technical axioms are necessary:
\begin{enumerate}[,label=\arabic*), leftmargin=5em]
\setcounter{enumii}{2}
\item The space $\cc(\pp,\RR)_\kk$ is closed under the Jordan product defined by equations \eqref{Jsq} and \eqref{Jdef}.
\item The pure state space of $\cc(\pp,\RR)_\kk$, seen as a Jordan-Lie-Banach algebra, coincides with $\pp$.
\end{enumerate}
See \cite[Theorem I.3.9.2. and Corollary I.3.9.2. pp 105--106]{landsman1998} for the details and proofs.\vspace*{0.1em}
}
\end{enumerate}
\end{theorem}

These are hence the two essential features that differentiate quantum kinematics from classical kinematics. As the name suggests, the first axiom---also called the ``two-sphere property"---is nothing but the geometric reformulation of the superposition principle\footnote{%
Indeed, in its core, the quantum superposition principle is a claim about the ability to generate new possible states from the knowledge of just a few: given the knowledge of states $p_1$ and $p_2$, one can deduce the existence of an infinite set $S_{p_1,p_2}$ of other states which are equally accessible to the system. In the standard Hilbert space formalism, the superposition principle is described by the canonical association of a two-dimensional complex vector space to any pair of states: for two different states $\psi_1, \psi_2 \in \hh$, any superposition of them can be written as $\phi = a \psi_1 + b \psi_2$, with $a,b \in \CC$. In other words, it is captured by the existence of a map
$$V: \hh \times \hh \longrightarrow \text{Hom}(\CC^2, \hh)$$
where the linear map $V_{\psi_1, \psi_2}: \CC^2 \rightarrow \hh$ is an injection iff $\psi_1$ and $\psi_2$ are linearly independent vectors. The geometric reformulation is then found simply by taking the projective analogue of this. Therein, the superposition principle is now seen as the existence of a map
$$S: \ph \times \ph \longrightarrow \text{Hom}(\PP\CC^2 \simeq S^2, \ph)\vspace*{-0.5em}%
$$
where, for $p_1 \neq p_2$, the map $S_{p_1,p_2}$ is an injection. Axiom \ref{PoS} is the generalization of this for any Poisson space with transition probability. 
}.
This has been invariably stressed, from the beginning of quantum mechanics, as one of the fundamental features of the theory. The second point, however, seems to have been the blind spot in the conceptual analysis of quantum kinematics.   

\section{Conclusion}
\label{conclusion}

In both classical and quantum kinematics, observables play two different conceptual roles: on the one hand, observables are \emph{quantities} which may take definite numerical values on certain states; on the other hand, observables are intimately related to the \emph{generation of transformations} on the space of states. The main goal of this work was to show that the detailed study of the two-fold role of observables furnishes a very fruitful point of view from which to compare the conceptual structure of classical and quantum kinematics. As our analysis puts forward, this double nature of observables is an essential feature of the theory, deeply related to the mathematical structures used in the description of classical and quantum kinematics. This is particularly salient in the geometric definition of observables (\autopageref{def:o}) and in Landmsan's axiomatization of quantum mechanics (\autopageref{PoS}). Yet, much more remains to be explored about the subject and its conceptual implications. In particular, it remains unclear how to interpret this two-fold role from a physical point of view. Why is it the case that observables are intimately related to the generation of transformations? And how, if at all, should we alter our conception of what observables are in order to render their two-fold role more natural? These are important questions which ought to receive more attention in the future.

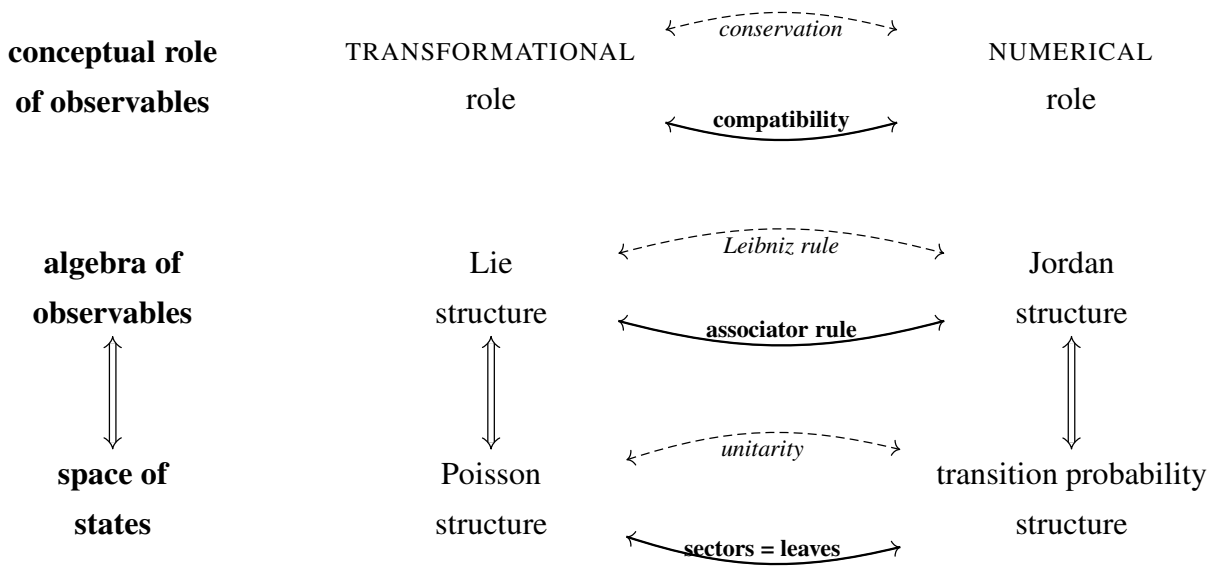
\begin{figure}[ht!]
\vspace*{-1em}
$$\begin{tikzcd}[]
\parbox{7em}{\centering \textbf{conceptual role \\ of observables}}
& \parbox{10em}{\centering \textsc{transformational}\\ role}  
\arrow[rrr, leftrightarrow, "\emph{conservation}"', bend left= 15, dashed] 
\arrow[rrr, leftrightarrow, "\textbf{compatibility}", bend right= 15]
\arrow[rrr, leftrightarrow, bend right= 15, dash, line width = 0.3mm] 
& 
& 
& \parbox{10em}{\centering \textsc{numerical} \\ role} 
\\ \\
\parbox{7em}{\centering \textbf{algebra of \\ observables}}  \ar[dd, Leftrightarrow]
& \parbox{7em}{\centering Lie \\ structure}  
\arrow[rrr, leftrightarrow, "\emph{Leibniz rule}"', bend left= 15, dashed] 
\arrow[rrr, leftrightarrow, "\textbf{associator rule}", bend right= 15] 
\arrow[rrr, leftrightarrow, bend right= 15, dash, line width = 0.3mm]
\ar[dd, Leftrightarrow]
& 
&
& \parbox{7em}{\centering Jordan \\ structure} \ar[dd, Leftrightarrow]
\\ \\
\parbox{7em}{\centering \textbf{space of \\ states}}  
& \parbox{7.5em}{\centering Poisson \\ structure}
\arrow[rrr, leftrightarrow, "\emph{unitarity}"', bend left= 16, dashed] 
\arrow[rrr, leftrightarrow, "\textbf{~~sectors = leaves~~}", bend right= 16]
\arrow[rrr, leftrightarrow, bend right= 16, dash, line width = 0.3mm] 
&
&
& \parbox{10em}{\centering transition probability \\ structure}	
\end{tikzcd}$$
\caption{Conceptual structure of kinematics. In both the Classical and the Quantum, observables play a two-fold role vis-\`a-vis of states which gets translated mathematically into the presence of two algebraic structures on the set of observables and two geometric structures on the space of states. The vertical double arrows definability equivalences: one can define the algebraic structures in terms of the geometric structures or vice versa. The dashed arrows represent a first compatibility condition, common to both kinematics. In contrast, the thick arrows represent the distinguishing feature of quantum kinematics: a compatibility condition which ties together the behaviour of the quantities (numerical role) and the transformations they generate.}
\label{fig:summary}
\end{figure}

To conclude, let us then summarize the global picture which has emerged from our analysis of three different formulations of quantum kinematics: the standard one in terms of Hilbert spaces, the geometric one in terms of K\"ahler manifolds and Landsman's in terms of Poisson spaces with a transition probability (see also \autoref{fig:summary}, \autopageref{fig:summary}).   

Common to both kinematics is the fact that the full description of observables is the conjunction of their numerical and transformational roles. That this two-fold role is the \emph{defining} feature of physical observables is best seen in their geometric definition: an observable is a function on the space of states to which an infinitesimal state transformation can be associated (cf. \autopageref{def:o}). Algebraically, this two-fold role gets translated into the existence of two structures on the set of observables: a Jordan product which governs the numerical role, and a Lie product which governs the transformational role. Accordingly, the language of real Jordan-Lie algebras is the common algebraic language which covers both classical and quantum kinematics. The geometric level of states mirrors the algebraic level in every respect: herein, the two-fold role manifests itself by the presence of two geometric structures---a transition probability structure and a Poisson structure (which respectively stem from the Jordan and Lie product, and from which the Jordan and Lie product can be defined)---and the common geometric language is that of Poisson spaces with a transition probability. One often restricts attention to the simpler case where the Poisson space has only one symplectic leaf. Then, the Poisson structure is equivalent to a symplectic 2-form and the non-trivial transition probability structure of the Quantum may be perceived as arising from a Riemannian metric (the transition probability being, roughly, the distance between two points). In this way, one recovers the geometric formulation of classical and quantum kinematics in terms of symplectic manifolds and K\"ahler manifolds respectively.

With the use of either Jordan-Lie algebras or Poisson spaces with a transition probability, one may sharply characterize the mathematical difference between the two kinematics. Contrary to the mistaken commutativity/non-commutativity motto, this point of view shows that, at the algebraic level, the difference really lies in the associativity/non-associativity of the Jordan product. Geometrically, the difference is captured in the trivia\-lity/non-triviality of the transition probability structure. In particular, this implies that, from the restricted point of view of the symplectic/Lie structure, the Classical and the Quantum are indistinguishable. In other words, both theories are identical if one focuses only on the transformational role of observables.

The conceptual difference between the Classical and the Quantum can only be grasped when studying the relation between the two roles of observables. In both kinematics, the transformations preserve the numerical role of the observables. At the geometric level, this fact is captured by unitarity: the Hamiltonian flow of any physical property preserves the transition probabilities. At the algebraic level, this is encoded in the Leibniz rule. However, on top of this, quantum kinematics exhibits a second compatibility condition between the two roles which distinguishes it from classical kinematics: the numerical role of observables encodes information on their transformational role. Geometrically, this is seen in the coincidence between the two natural foliations on the space of states induced by the two geometric structures. Algebraically, it is encoded in the associator rule, which ties together the Jordan and Lie structures.

In the light of Landsman's axiomatization of quantum mechanics, we see that this last point may be turned around: given the two-fold role of observables in kinematics, the requirement that the two roles be consistent with each other forces the  Jordan product to be non-associative and the transition probability to be non-trivial. Therefore, \emph{the two-fold role compatibility condition and the superposition principle may be seen as the two fundamental pillars on which quantum kinematics rests}.

\vspace*{1.5em}
\noindent \textbf{Acknowledgements.} I would like to thank Gabriel Catren, Mathieu Anel, Christine Cachot, Julien Page, Michael Wright, Fernando Zalamea and anonymous reviewers for helpful discussions and comments on earlier drafts of this paper.
\vspace*{1em}
{\small
\bibliographystyle{spmpsci}      
\bibliography{./TwoFold_Role.bib}} 

\begin{thebibliography}{10}
\providecommand{\url}[1]{{#1}}
\providecommand{\urlprefix}{URL }
\expandafter\ifx\csname urlstyle\endcsname\relax
  \providecommand{\doi}[1]{DOI~\discretionary{}{}{}#1}\else
  \providecommand{\doi}{DOI~\discretionary{}{}{}\begingroup
  \urlstyle{rm}\Url}\fi

\bibitem{abraham1978}
Abraham, R., Marsden, J.E.: Foundations of {M}echanics, 2nd edn.
\newblock Addison-Wesley Publishing Company, Redwood City (1978)

\bibitem{abraham1988}
Abraham, R., Marsden, J.E., Ratiu, T.S.: Manifold, Tensor Analysis, and
  Applications, 2nd edn.
\newblock Springer-Verlag, New York (1988)

\bibitem{alfsen1976}
Alfsen, E.M., Shultz, F.W.: Non-commutative spectral theory for affine function
  spaces on convex sets.
\newblock Memoirs of the American Mathematical Society \textbf{172}, 1--120
  (1976)

\bibitem{alfsen2001}
Alfsen, E.M., Shultz, F.W.: State Spaces of Operator Algebras.
\newblock Birkh{\"a}user, Boston (2001)

\bibitem{arnold1989}
Arnold, V.I.: Mathematical Methods of Classical Mechanics, vol.~60, 2nd edn.
\newblock Springer-Verlag, New York (1989)

\bibitem{ashtekar1997}
Ashtekar, A., Schilling, T.A.: Geometrical formulation of quantum mechanics.
\newblock In: A.~Harvey (ed.) On Einstein's Path: {E}ssays in {H}onor of
  {E}ngelbert {S}ch\"ucking, pp. 23--65. Springer, New York (1997)

\bibitem{bargmann1954}
Bargmann, V.: On unitary ray representations of continuous groups.
\newblock Annals of Mathematics \textbf{59}(1), 1--46 (1954)

\bibitem{bell1964}
Bell, J.S.: On the {E}instein-{P}odolsky-{R}osen paradox.
\newblock Physics \textbf{1}(3), 195--200 (1964)

\bibitem{born1967a}
Born, M., Heisenberg, W., Jordan, P.: On quantum mechanics {II}.
\newblock In: B.~Van~der Waerden (ed.) Sources of Quantum Mechanics, pp.
  321--384. Dover Publications, Inc., New York (1967)

\bibitem{brody2001}
Brody, D.C., Hughston, L.P.: Geometric quantum mechanics.
\newblock Journal of Geometry and Physics \textbf{38}(1), 19--53 (2001)

\bibitem{cartier2008}
Cartier, P.: Notion de spectre.
\newblock In: Premi{\`e}re {\'e}cole d'{\'e}t{\'e} : Histoire conceptuelle des
  math{\'e}matiques - Dualit{\'e} Alg{\`e}bre-G{\'e}om{\'e}trie, pp. 232--242.
  Maison des Sciences de l'Homme, Universidade de Brasilia (2008)

\bibitem{catren2008}
Catren, G.: On classical and quantum objectivity.
\newblock Foundations of Physics \textbf{38}(5), 470--487 (2008)

\bibitem{catren2014a}
Catren, G.: Quantum ontology in the light of gauge theories.
\newblock In: S.A. C.~de Ronde, D.~Aerts (eds.) Probing the {M}eaning of
  {Q}uantum {M}echanics: {P}hysical, {P}hilosophical, and {L}ogical
  {P}erspectives. World Scientific (2014)

\bibitem{chernoff1974}
Chernoff, P.R., Marsden, J.E.: Properties of Infinite Dimensional Hamiltonian
  Systems.
\newblock Lecture Notes in Mathematics. Springer-Verlag, Heidelberg (1974)

\bibitem{cirelli1999}
Cirelli, R., Gatti, M., Mani{\`a}, A.: On the nonlinear extension of quantum
  superposition and uncertainty principles.
\newblock Journal of Geometry and Physics \textbf{29}(1), 64--86 (1999)

\bibitem{cirelli2003}
Cirelli, R., Gatti, M., Mani{\`a}, A.: The pure state space of quantum
  mechanics as hermitian symmetric space.
\newblock Journal of Geometry and Physics \textbf{45}(3), 267--284 (2003)

\bibitem{dirac1925}
Dirac, P.A.M.: The fundamental equations of quantum mechanics.
\newblock Proceedings of the Royal Society of London \textbf{A109}, 642--653
  (1925)

\bibitem{dirac1958}
Dirac, P.A.M.: The Principles of Quantum Mechanics, 4th ed. edn.
\newblock Oxford University Press, Oxford (1958)

\bibitem{gelfand1943}
Gelfand, I., Naimark, M.: On the imbedding of normed rings into the ring of
  operators in {H}ilbert space.
\newblock Matematicheskii Sbornik \textbf{12}, 197--213 (1943)

\bibitem{geroch1971}
Geroch, R.: A method for generating solutions of {E}instein's equations.
\newblock Journal of Mathematical Physics \textbf{12}(6), 918--924 (1971)

\bibitem{gleason1957}
Gleason, A.M.: Measures on the closed subspaces of a {H}ilbert space.
\newblock Journal of Mathematics and Mechanics \textbf{6}, 885--894 (1957)

\bibitem{grgin1974}
Grgin, E., Petersen, A.: {Duality of observables and generators in classical
  and quantum mechanics}.
\newblock Journal of Mathematical Physics \textbf{15}(6), 764--769 (1974)

\bibitem{guillemin2006}
Guillemin, V., Sternberg, S.: Variations on a Theme by Kepler, vol.~42.
\newblock American Mathematical Soc. (2006)

\bibitem{heisenberg1925}
Heisenberg, W.: {\"Uber quantentheoretische Umdeutung kinematischer und
  mechanischer Beziehungen}.
\newblock Zeitschrift f{\"u}r Physik \textbf{33}, 879--893 (1925)

\bibitem{iglesias2014}
Iglesias-Zemmour, P.: Aper{\c c}u des origines de la g{\'e}om{\'e}trie
  symplectique.
\newblock In: Actes du colloque ``Histoire des g{\'e}om{\'e}tries'', vol. 1,
  Maison des Sciences de l'Homme. Paris (2004)

\bibitem{jordan1933}
Jordan, P.: {\"Uber Verallgemeinerungsm\"oglichkeiten des Formalismus der
  Quantenmechanik}.
\newblock Nachr. Akad. Wiss. G\"ottingen \textbf{41}, 209--217 (1933)

\bibitem{kibble1979}
Kibble, T.: Geometrization of quantum mechanics.
\newblock Communications in Mathematical Physics \textbf{65}(2), 189--201
  (1979)

\bibitem{kobayashi1969}
Kobayashi, S., Nomizu, K.: Foundations of Differential Geometry, vol.~2.
\newblock Wiley, New York (1969)

\bibitem{kochen1967}
Kochen, S.B., Specker, E.: The problem of hidden variables and physical
  measurements.
\newblock Journal of Mathematics and Mechanics \textbf{17}(59--67) (1967)

\bibitem{kosmann2011}
Kosmann-Schwarzbach, Y.: The Noether Theorems: Invariance and Conservations
  Laws in the Twentieth Century.
\newblock Sources and Studies in the History of Mathematics and Physical
  Sciences. Springer, New York (2011)

\bibitem{kostant1970}
Kostant, B.: Quantization and unitary representations.
\newblock In: Lectures in Modern Analysis and Applications III, Lecture Notes
  in Mathematics, pp. 87--208. Springer-Verlag (1970)

\bibitem{landsman1994}
Landsman, N.P.: Classical and quantum representation theory.
\newblock arXiv preprint  (1994).
\newblock \urlprefix\url{http://arxiv.org/abs/hep-th/9411172}

\bibitem{landsman1997}
Landsman, N.P.: Poisson spaces with a transition probability.
\newblock Review of Mathematical Physics \textbf{9}(1), 29--57 (1997)

\bibitem{landsman1998}
Landsman, N.P.: Mathematical Topics Between Classical and Quantum Mechanics.
\newblock Springer, New York (1998)

\bibitem{landsman1998c}
Landsman, N.P.: Simple new axioms for quantum mechanics.
\newblock International Journal of Theoretical Physics \textbf{37}, 343--348
  (1998)

\bibitem{marsden1999}
Marsden, J.E., Ratiu, T.S.: Introduction to Mechanics and Symmetry. A Basic
  Exposition of Classical Mechanical Systems, 2nd ed. edn.
\newblock Springer, New York (1999)

\bibitem{marsden1974}
Marsden, J.E., Weinstein, A.: Reduction of symplectic manifolds with symmetry.
\newblock Reports on Mathematical Physics \textbf{5}(1), 121--130 (1974)

\bibitem{neumann1955}
von Neumann, J.: Mathematical Foundations of Quantum Mechanics.
\newblock Princeton University Press, Princeton (1955)

\bibitem{neumann1981}
von Neumann, J.: Continuous Geometries with a Transition Probability, vol. 252.
\newblock American Mathematical Society (1981)

\bibitem{puta1993}
Puta, M.: Hamiltonian Mechanical Systems and Geometric Quantization.
\newblock Kluwer Academic Publishers, Dordrecht, The Netherlands (1993)

\bibitem{schilling1996}
Schilling, T.A.: Geometry of {Q}uantum {M}echanics.
\newblock Ph.D. thesis, The Pennsylvania State University (1996)

\bibitem{shultz1982}
Shultz, F.W.: Pure states as dual objects for {$C^*$}-algebras.
\newblock Communications in Mathematical Physics \textbf{82}, 497--509 (1982)

\bibitem{souriau1970}
Souriau, J.M.: Structure des syst{\`e}mes dynamiques.
\newblock Dunod, Paris (1970)

\bibitem{strocchi2008}
Strocchi, F.: An {I}ntroduction to the {M}athematical {S}tructure of {Q}uantum
  {M}echanics, 2nd edn.
\newblock World Scientific, Singapore (2008)

\bibitem{townsend2000}
Townsend, J.S.: A Modern Approach to Quantum Mechanics.
\newblock University Science Books, Sausalito (2000)

\bibitem{wigner1959}
Wigner, E.P.: Group Theory and Its Applications to the Quantum Mechanics of
  Atomic Spectra.
\newblock Academic Press, New York (1959)

\bibitem{woodhouse1991}
Woodhouse, N.: Geometric Quantization, 2nd edn.
\newblock Clarendon Press, Oxford (1991)

\end{thebibliography}
\end{document}